\pdfoutput=1
 
\documentclass[cernpreprint, atlasdraft=false, texlive=2023, UKenglish, texmf, orcidlogo]{atlasdoc}

\usepackage{atlaspackage}
\usepackage{atlasbiblatex}
 
\usepackage{atlasphysics}

\graphicspath{{logos/}{figures/}}

\usepackage{ANA-HDBS-2018-57-PAPER-defs}
\usepackage{multirow}
\usepackage{multicol}
\usepackage{graphicx}
\usepackage{subfig}
\usepackage{comment}
\usepackage{tabularx}
\usepackage{tikz}
\usepackage[compat=1.1.0]{tikz-feynman}
\usepackage{todonotes}

\AtlasTitle{Search for a new $Z'$ gauge boson in $4\mu$ events with the ATLAS experiment}

\AtlasAbstract{
This paper presents a search for a new $Z'$ vector gauge boson with the ATLAS experiment at the Large Hadron Collider using $pp$ collision data collected at $\sqrt{s} = 13$ TeV, corresponding to an integrated luminosity of 139 fb$^{-1}$.
The new gauge boson $Z'$ is predicted by $L_{\mu}-L_{\tau}$ models
to address observed phenomena that can not be explained by the Standard Model.
The search examines the four-muon (4$\mu$) final state,
using a deep learning neural network classifier to separate the $Z'$ signal from the Standard Model background events.
The di-muon invariant masses in the $4\mu$ events are used to extract the $Z'$ resonance signature.
No significant excess of events is observed over the predicted background.
Upper limits at a 95\% confidence level
on the $Z'$ production cross-section times the decay branching fraction of $pp \rightarrow Z'\mu\mu \rightarrow 4\mu$ are set from
0.31 to 4.3 fb
for the $Z'$ mass ranging from 5 to 81 GeV.
The corresponding
common coupling strengths, $g_{Z'}$, of the $Z'$ boson to the second and third generation leptons above
0.003 -- 0.2
have been excluded.
}

\AtlasRefCode{HDBS-2018-57}
 
\PreprintIdNumber{CERN-EP-2022-251}

\AtlasJournalRef{JHEP\ 07\ (2023)\ 90}
\AtlasDOI{DOI:10.1007/JHEP07(2023)090}

\hypersetup{pdftitle={ATLAS document},pdfauthor={The ATLAS Collaboration}}
 
\begin{document}
 
\maketitle

\clearpage

\section{Introduction}
\label{sec:intro}

This paper presents a search for a new vector boson $Z'$ in the four-muon (4$\mu$) final state with data recorded in proton-proton ($pp$) collisions at $\sqrt s$ = 13~\TeV{} by the ATLAS detector~\cite{PERF-2007-01} at the Large Hadron Collider (LHC), corresponding to an integrated luminosity of 139~fb$^{-1}$.
The new gauge boson $Z'$, which only interacts with the second and third generation leptons, is predicted by $L_{\mu}-L_{\tau}$ models~\cite{PhysRevD.44.2118} which extend the Standard Model (SM) with an additional $U(1)_{L_{\mu}-L_{\tau}}$ symmetry
to address the observed anomalies of the muon magnetic dipole moment ($g-2$)~\cite{Fermilab:Muong2,Keshavarzi_2019,Venanzoni:2016hP,PhysRevD.73.072003} and of rare B decays~\cite{Aaij2022,Aaij_2013,Aaij_2014,PhysRevLett.125.011802}.
These models also aim to probe outstanding physics questions related to dark matter and neutrino mass~\cite{Altmannshofer:2016jzy,Ma_2002,Langacker_2009}.
 
The \Zp kinematics, mass, and interactions (with the same coupling strength to the second and third lepton families), are described by the Lagrangian below:
\begin{equation*}
L_{Z'} = - {1\over 4} F_{\alpha\beta}F^{\alpha\beta} + {1\over 2} M^2_{Z'} Z'^\alpha Z'_\alpha
- g_{Z'} Z'_\alpha(\bar\ell_2\gamma^\alpha\ell_2 + \bar\mu\gamma^\alpha\mu
-\bar\ell_3\gamma^\alpha\ell_3 - \bar\tau\gamma^\alpha\tau ),
\end{equation*}
where $F_{\alpha\beta} = \partial_\alpha Z'_\beta - \partial_\beta Z'_\alpha$ is the $Z'$ field strength tensor;  $\ell_i = (\nu_i, e_i)^T$ ($i=2,3$, denoting the second and the third generation left-handed lepton doublets); $\mu$ and $\tau$ represent the right-handed muon and tau singlets; and $g_{Z'}$ (which will be referred to as \coupzp in the rest of this paper) is the interaction coupling constant.
The parameter space of ($m_{Z'}, \coupzp$) has not been strongly constrained in experiments since the $Z'$ does not directly couple to the electron nor to any quarks, hence it could not be directly produced from $e^+e^-$ and $pp$ colliding beams.

\begin{figure}[!ht]
\vspace{-0.6cm}
\centering
\subfloat[]{
\includegraphics[width=0.24\textwidth]{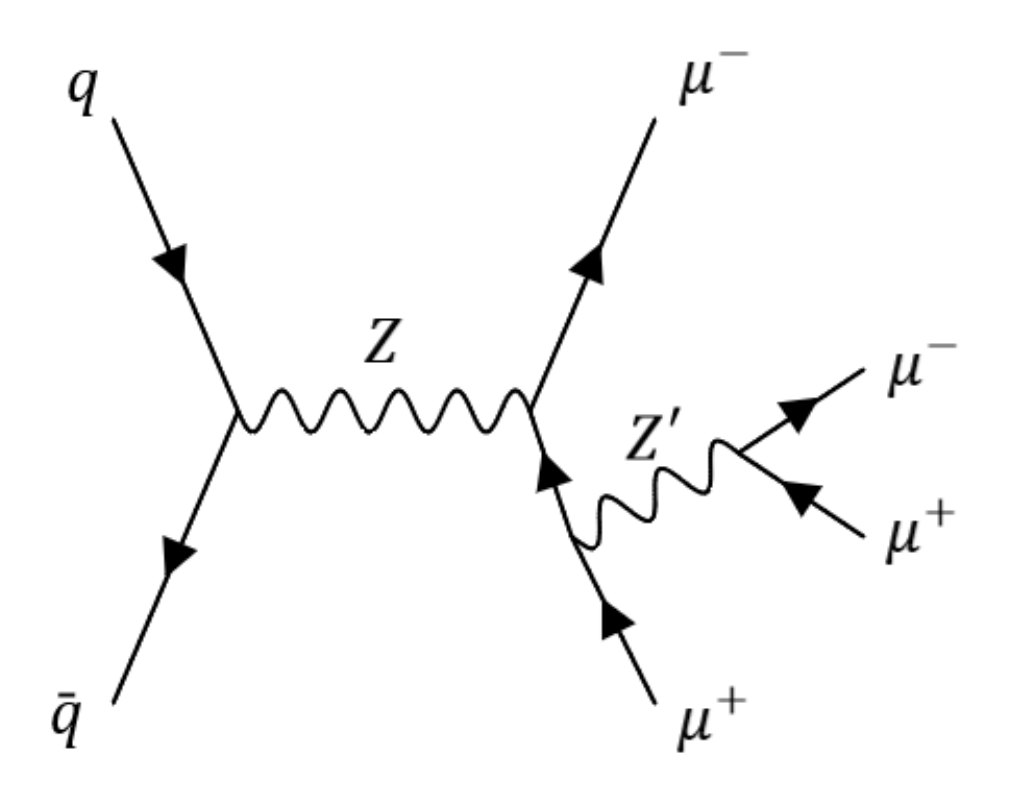}
\label{fig:feynman_new:a}
}
\subfloat[]{
\includegraphics[width=0.24\textwidth]{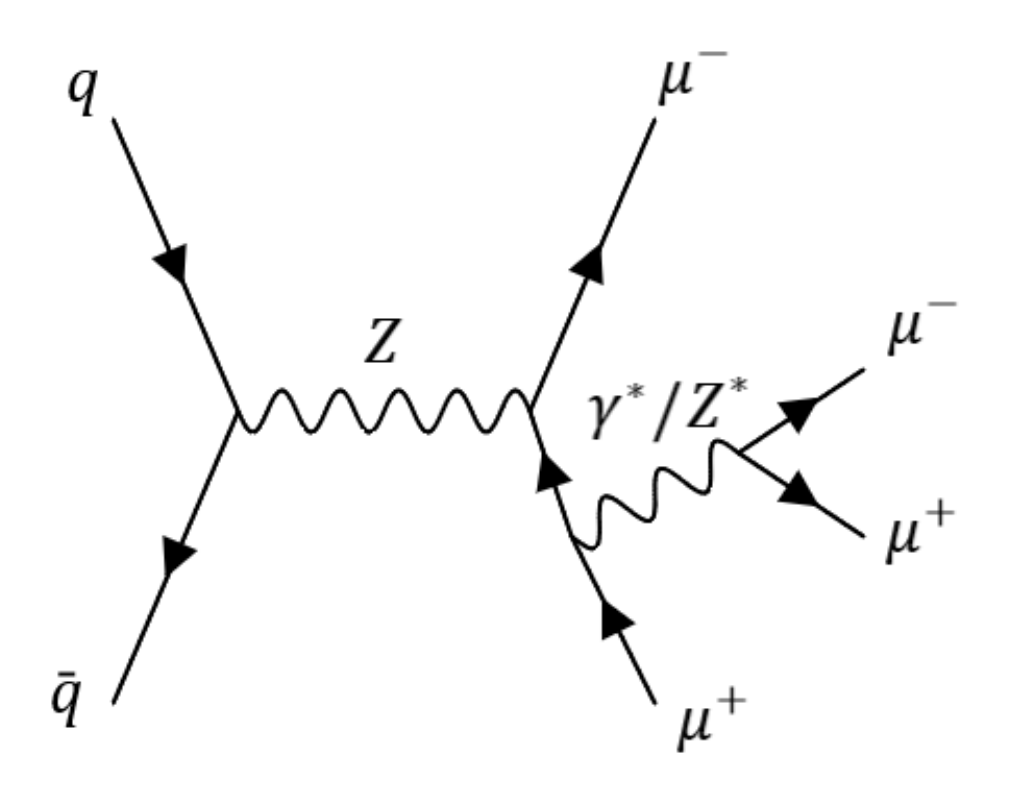}
\label{fig:feynman_new:b}
}
\subfloat[]{
\includegraphics[width=0.22\textwidth]{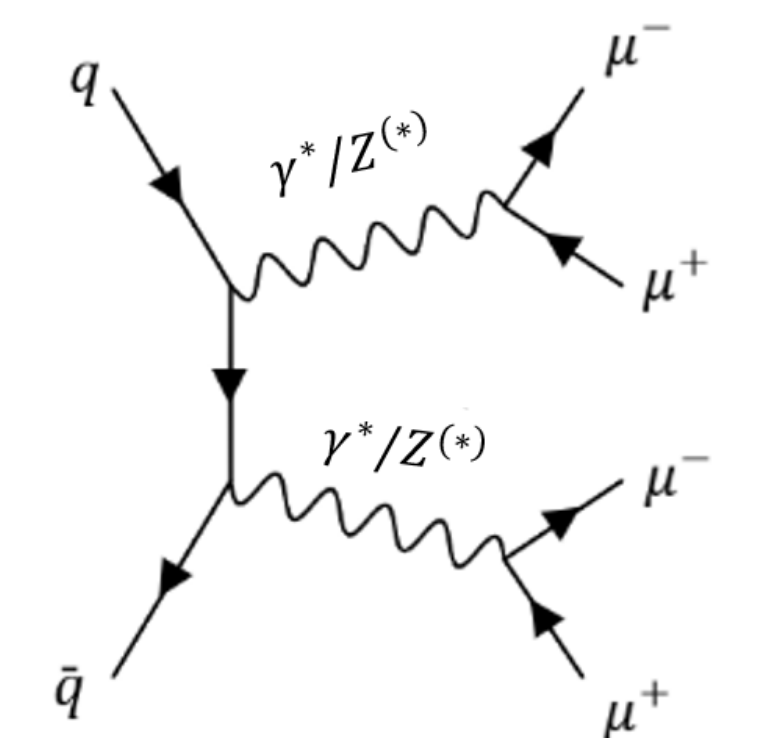}
\label{fig:feynman_new:c}
}
\subfloat[]{
\includegraphics[width=0.24\textwidth]{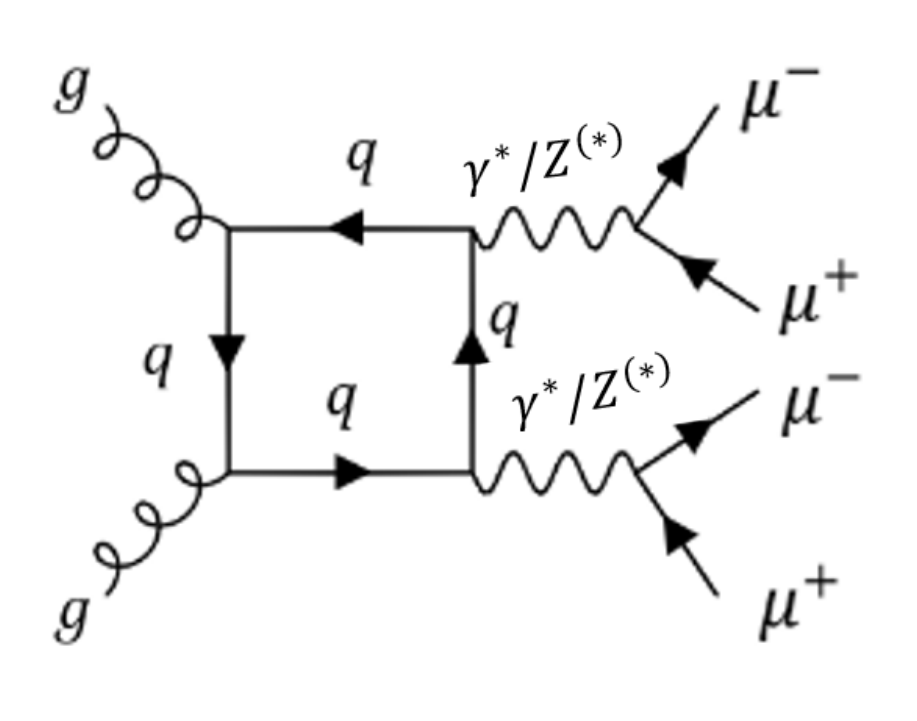}
\label{fig:feynman_new:d}
}
\caption{Feynman diagrams of $Z'$ production through radiation in a Drell-Yan process (a), and of the corresponding SM background processes (b - d) with a 4$\mu$ final state.}
\label{fig:feynman_new}
\end{figure}

In $pp$ collisions at the LHC, the $Z'$ could be produced from final state radiation of $\mu$ or $\tau$ pairs of the Drell-Yan (DY) process as shown in \Fig{\ref{fig:feynman_new:a}} with a $4\mu$ final state, which provides the cleanest signature to search for the $Z'$.
For relatively low $Z'$ mass, the most promising experimental signature would be an excess of 4$\mu$ events with the invariant mass of one $\mu^+\mu^-$ pair peaking around the $Z'$ mass.
The major background comes from the SM 4$\mu$ production processes shown in \Fig{\labelcref{fig:feynman_new:b,fig:feynman_new:c,fig:feynman_new:d}}.
The \Zp could also be produced in \Wboson production through the DY process, $pp\rightarrow W \rightarrow Z' \mu \nu \rightarrow 3\mu + \nu$.
The experimental signature would have a final state of 3$\mu$ plus large missing transverse energy.
This final state is not included in this analysis.
 
Both the ATLAS and CMS Collaborations have measured the cross-sections of the SM $Z \rightarrow 4\mu$ process~\cite{CMS:2012bw,Aad:2014wra,Sirunyan:2017zjc}.
The measurement by ATLAS was used by theorists to set limits in the parameter space of the $L_\mu - L_\tau$ model~\cite{Altmannshofer:2016jzy}.
The CMS Collaboration has directly searched for the $Z'$ boson in the mass region between 5 to 70~\GeV{} with the 4$\mu$ final state using 77.3~\ifb~of data~\cite{CMS-EXO-18-008}, and set upper limits on the $Z'$ to muon coupling strength, \coupzp, of 0.004 -- 0.3 at 95\% confidence level, depending on the $Z'$ mass.
 
The organization of this paper is as follows.
The ATLAS detector is described in the next section.
The dataset and Monte Carlo samples used in this analysis are detailed in \Sect{\ref{sec:data-MC}}.
Event reconstruction and selection, followed by background estimation from data, are described in \Sect{\ref{sec:simu}} and \Sect{\ref{sec:background}}.
Event classification using a deep learning approach is presented in \Sect{\ref{sec:dnn}}.
Systematic uncertainties and the statistical approach to interpret data to obtain the 
results
are reported in \Sect{\ref{sec:syst}} and \Sect{\ref{sec:result}}.
The conclusion is given in the final section.


\section{ATLAS Detector}
\label{sec:detector}

 
\newcommand{\AtlasCoordFootnote}{
ATLAS uses a right-handed coordinate system with its origin at the nominal interaction point (IP)
in the centre of the detector and the \(z\)-axis along the beam pipe.
The \(x\)-axis points from the IP to the centre of the LHC ring,
and the \(y\)-axis points upwards.
Cylindrical coordinates \((r,\phi)\) are used in the transverse plane,
\(\phi\) being the azimuthal angle around the \(z\)-axis.
The pseudorapidity is defined in terms of the polar angle \(\theta\) as \(\eta = -\ln \tan(\theta/2)\).
Angular distance is measured in units of \(\Delta R \equiv \sqrt{(\Delta\eta)^{2} + (\Delta\phi)^{2}}\).}
 
The ATLAS detector~\cite{PERF-2007-01} at the LHC covers nearly the entire solid angle around the collision point.\footnote{\AtlasCoordFootnote}
It consists of an inner tracking detector surrounded by a thin superconducting solenoid, electromagnetic and hadron calorimeters,
and a muon spectrometer incorporating three large superconducting air-core toroidal magnets.
 
The inner-detector system (ID) is immersed in a \SI{2}{\tesla} axial magnetic field
and provides charged-particle tracking in the range \(|\eta| < 2.5\).
The high-granularity silicon pixel detector covers the vertex region and typically provides four measurements per track,
the first hit normally being in the insertable B-layer (IBL) installed before Run~2~\cite{ATLAS-TDR-19,PIX-2018-001}.
It is followed by the silicon microstrip tracker (SCT), which usually provides eight measurements per track.
These silicon detectors are complemented by the transition radiation tracker (TRT),
which enables radially extended track reconstruction up to \(|\eta| = 2.0\).
The TRT also provides electron identification information
based on the fraction of hits (typically 30 in total) above a higher energy-deposit threshold corresponding to transition radiation.
 
The calorimeter system covers the pseudorapidity range \(|\eta| < 4.9\).
Within the region \(|\eta|< 3.2\), electromagnetic calorimetry is provided by barrel and
endcap high-granularity lead/liquid-argon (LAr) calorimeters,
with an additional thin LAr presampler covering \(|\eta| < 1.8\)
to correct for energy loss in material upstream of the calorimeters.
Hadron calorimetry is provided by the steel/scintillator-tile calorimeter,
segmented into three barrel structures within \(|\eta| < 1.7\), and two copper/LAr hadron endcap calorimeters.
The solid angle coverage is completed with forward copper/LAr and tungsten/LAr calorimeter modules
optimised for electromagnetic and hadronic energy measurements respectively.
 
The muon spectrometer (MS) comprises separate trigger and
high-precision tracking chambers measuring the deflection of muons in a magnetic field generated by the superconducting air-core toroidal magnets.
The field integral of the toroids ranges between \num{2.0} and \SI{6.0}{\tesla\metre}
across most of the detector.
Three layers of precision chambers, each consisting of layers of monitored drift tubes, covers the region \(|\eta| < 2.7\),
complemented by cathode-strip chambers in the forward region, where the background is highest.
The muon trigger system covers the range \(|\eta| < 2.4\) with resistive-plate chambers in the barrel, and thin-gap chambers in the endcap regions.
 
Interesting events are selected by the first-level trigger system implemented in custom hardware,
followed by selections made by algorithms implemented in software in the high-level trigger~\cite{TRIG-2016-01}.
The first-level trigger accepts events from the \SI{40}{\MHz} bunch crossings at a rate below \SI{100}{\kHz},
which the high-level trigger further reduces in order to record events to disk at about \SI{1}{\kHz}.
 
An extensive software suite~\cite{ATL-SOFT-PUB-2021-001} is used in data simulation, in the reconstruction
and analysis of real and simulated data, in detector operations, and in the trigger and data acquisition
systems of the experiment.


\section{Dataset and Monte Carlo Simulations}
\label{sec:data-MC}

The data used for this analysis were recorded using single-muon and multi-muon triggers, corresponding to an integrated luminosity of 139~\ifb~after the application of data quality requirements~\cite{Aad_2020}.
The transverse momentum (\pT) thresholds for the single-muon trigger vary from 20 to 26 GeV, for the di-muon trigger from 10 to 14 GeV, and for the tri-muon trigger from 4 to 6 GeV, depending on the data-taking periods~\cite{TRIG-2018-01}.
The overall trigger efficiency in the phase space defined by the event selections used in this analysis is higher than 98\%.

\subsection{Simulation of \texorpdfstring{$Z'$}{Z'} Production}
\label{sec:simulation_sig}
 
The $L_\mu - L_\tau$ model is used for Monte Carlo (MC) \Zp signal sample production, where the \Zp couples to the left-handed (LH) muon or tau leptons and their corresponding neutrinos, and to the right-handed (RH) muon or tau leptons.
In the model, the \Zp couplings to the first lepton families (electron and its neutrino) and all quarks are set to zero.
The branching fractions of the \Zp decay to a pair of muons and a pair of muon neutrinos are
$\frac{1}{3}$ and $\frac{1}{6}$, respectively.
The signal from tau decays in the 4$\mu$ final state is found to be negligible in this analysis and is not included as signal.
 
The interactions~\cite{delAguila:1972795} mediated by a resonance $Z'$ which couples to the second and third generation leptons are used for four-muon signal event generation.
The signal events are generated with \MGNLO[2.7.3]~\cite{Alwall:2014hca} at leading-order (LO) accuracy in QCD  by using the \textit{Universal FeynRules Output} (UFO) format~\cite{Degrande:2011ua,Alloul:2013bka}.
Based on theoretical calculations~\cite{Grazzini_2015, h2020resummed} for related processes,
the appropriate NNLO/LO correction factor K of 1.3 is used to correct the MC LO signal cross-sections.
In the signal production simulations, the \NNPDF[2.3nlo] set~\cite{Ball:2012cx} is used for parton distribution function (PDF) for the $pp$ collisions.
\PHOTOSpp[3.61]~\cite{Golonka:2005pn, Davidson:2010ew} is used to simulate the effect of QED final-state radiation.
 
The MC simulated events are generated for a range of masses and coupling parameters of the $L_\mu-L_\tau$ model.
The $Z'$ mass ranges from 5 to 81 GeV, and the value of coupling constant \coupzp ranges from 0.008 to 0.316, as summarized in Table~\ref{tab:signalSamples}.
The value of the \coupzp at each \Zp mass was chosen to be close to the expected experimental sensitivity to allow a sensitive search for a very small \Zp signal in the full Run 2 dataset.

With the chosen \coupzp, the natural width $\Gamma$ of the $Z'$
and the cross-section, both are proportional to $g^2$,
are calculated as listed in \Tab{\ref{tab:signalSamples}}.
It should be noted that all the generated mass points have the ratio $\Gamma / m_{Z'}$ much smaller than 1\%.
Therefore, the ATLAS detector mass resolution (2\%)  dominates the width of the signal Z' mass peak in the experimental dimuon mass spectrum.
 
\begin{table}[h!]
\caption{
Summary of the chosen $Z'$ hypotheses and corresponding coupling, width, and cross-section\\
$\sigma$ ($pp\rightarrow Z' \mu^+\mu^- \rightarrow \mu^+\mu^-\mu^+\mu^-$), calculated at LO accuracy in QCD, at each mass point.
}
\centering
\resizebox{0.9\textwidth}{!}{
\begin{tabular}{cccc|cccc}
\hline
$m_{Z'}$ {[}GeV{]} & \coupzp & $\Gamma$ {[}GeV{]}        & $\sigma$ {[}fb{]} & $m_{Z'}$ {[}GeV{]} & \coupzp & $\Gamma$ {[}GeV{]} & $\sigma$ {[}fb{]} \\ \hline
5                  & 0.0080  & \SI{2.45E-5}{}    \SI{}{} & 9.96              & 42                 & 0.0900  & \SI{2.71E-2}{}     & 13.38             \\
7                  & 0.0085  & \SI{3.99E-5}{}            & 7.06              & 45                 & 0.1000  & \SI{3.58E-2}{}     & 11.72             \\
9                  & 0.0090  & \SI{5.78E-5}{}            & 5.60              & 48                 & 0.1100  & \SI{4.62E-2}{}     & 9.96              \\
11                 & 0.0095  & \SI{7.89E-5}{}            & 4.65              & 51                 & 0.1200  & \SI{5.84E-2}{}     & 8.24              \\
13                 & 0.0100  & \SI{1.03E-4}{}            & 3.95              & 54                 & 0.1600  & \SI{1.10E-1}{}     & 10.07             \\
15                 & 0.0120  & \SI{1.72E-4}{}            & 4.45              & 57                 & 0.2000  & \SI{1.81E-1}{}     & 10.73             \\
17                 & 0.0140  & \SI{2.65E-4}{}            & 4.80              & 60                 & 0.2665  & \SI{3.39E-1}{}     & 12.92             \\
19                 & 0.0160  & \SI{3.87E-4}{}            & 5.00              & 63                 & 0.2680  & \SI{3.60E-1}{}     & 8.84              \\
23                 & 0.0240  & \SI{1.05E-3}{}            & 7.30              & 66                 & 0.2780  & \SI{4.06E-1}{}     & 6.50              \\
27                 & 0.0320  & \SI{2.20E-3}{}            & 8.50              & 69                 & 0.2890  & \SI{4.59E-1}{}     & 4.89              \\
31                 & 0.0400  & \SI{3.95E-3}{}            & 8.72              & 72                 & 0.3000  & \SI{5.15E-1}{}     & 3.80              \\
35                 & 0.0600  & \SI{1.00E-2}{}            & 12.82             & 75                 & 0.3000  & \SI{5.37E-1}{}     & 2.88              \\
39                 & 0.0800  & \SI{1.99E-2}{}            & 14.77             & 78                 & 0.3080  & \SI{5.89E-1}{}     & 2.40              \\
&         &                           &                   & 81                 & 0.3160  & \SI{6.44E-1}{}     & 2.08              \\
\hline
\end{tabular}
}
\label{tab:signalSamples}
\end{table}
 
The $Z'$ signal samples were simulated with the ATLAS fast simulation framework (Atlfast-II)~\cite{SOFT-2010-01}
to produce predictions that can be directly compared with the data.
 
\subsection{Simulation of Background Events}
\label{sec:simulation_bkg}
 
Dominant SM backgrounds in this analysis come from the SM $Z \rightarrow 4\mu$ processes where the four leptons have an invariant mass close to that of the $Z$ boson.
In the higher mass region the $ZZ^*$ production contributes a sizable number of prompt $4\mu$ events.
In addition, there are very small contributions from the Higgs boson, $t\bar{t}V$ ($V=W,Z$), and tri-boson ($VVV$) production processes.
These events are estimated with MC simulations.
The background events including muons from heavy-flavour hadron decays, misidentified jets, or photon conversions (collectively referred to as “non-prompt muon background") are mostly coming from \Zjet, \ttbar and single-top-quark production processes and estimated from data in this analysis, as described in \Sect{\ref{sec:background}}.
The \Zjet and \ttbar MC samples are also produced for background studies.
 
Samples of diboson production $q\bar q\rightarrow VV^{(*)}$,
including the processes shown in Figures \ref{fig:feynman_new:b} and \ref{fig:feynman_new:c}, were simulated with the \SHERPA[2.2.2]~\cite{Bothmann:2019yzt} generator, including off-shell effects and Higgs boson contributions, where appropriate.
Fully leptonic final states and semileptonic final states, where one boson decays leptonically and the other hadronically, were generated using matrix elements at next-to-leading-order (NLO) accuracy in QCD for up to one additional parton and at LO accuracy for up to three additional parton emissions.
Samples for the loop-induced processes
$gg \to ZZ^{(*)}$
, shown in Figure~\ref{fig:feynman_new:d}, were generated using LO-accurate matrix elements for up to one additional parton emission for both the cases of fully leptonic and semileptonic final states.
The matrix element calculations were matched and merged with the \SHERPA parton shower based on Catani--Seymour dipole factorisation~\cite{Gleisberg:2008fv,Schumann:2007mg} using the MEPS@NLO prescription~\cite{Hoeche:2011fd,Hoeche:2012yf,Catani:2001cc,Hoeche:2009rj}.
The virtual QCD corrections were provided by the \OPENLOOPS library~\cite{Buccioni:2019sur,Cascioli:2011va,Denner:2016kdg}.
The \NNPDF[3.0nnlo] set of PDFs was used~\cite{Ball:2014uwa}, along with the
dedicated set of parton-shower parameters (tune) developed by the
\SHERPA authors.
 
The production of \ttbar events was modelled using the \POWHEGBOX[v2]~\cite{Frixione:2007nw,Nason:2004rx,Frixione:2007vw,Alioli:2010xd} generator at NLO in QCD with the \NNPDF[3.0nlo] PDF set and the \hdamp parameter\footnote{The \hdamp parameter is a resummation damping factor and one of the parameters that controls the matching of \POWHEG matrix elements to the parton shower and thus effectively regulates the high-\pt radiation against which the \ttbar system recoils.} set to 1.5 \mtop~\cite{ATL-PHYS-PUB-2016-020}.
The events were interfaced to \PYTHIA[8.230]~\cite{Sjostrand:2014zea} to model the parton shower, hadronisation, and underlying event, with parameters set according to the A14 tune~\cite{ATL-PHYS-PUB-2014-021} and using the \NNPDF[2.3lo] set of PDFs.
The associated production of top quarks with $W$ bosons ($tW$) was modelled by the \POWHEGBOX[v2]~\cite{Re:2010bp,Nason:2004rx,Frixione:2007vw,Alioli:2010xd} generator at NLO in QCD using the five-flavour scheme and the \NNPDF[3.0nlo] set of PDFs.
The diagram removal scheme~\cite{Frixione:2008yi} was used to remove interference and overlap with \ttbar production.
The events were interfaced to \PYTHIA[8.230] using the A14 tune and the \NNPDF[2.3lo] set of PDFs.
 
The production of
$V+$jets
was simulated with the \SHERPA[2.2.1] generator using NLO matrix elements for up to two partons, and LO matrix elements for up to four partons, calculated with the Comix and \OPENLOOPS libraries.
They were matched with the \SHERPA parton shower using the \MEPSatNLO prescription with the set of tuned parameters developed by the \SHERPA authors.
The \NNPDF[3.0nnlo] set of PDFs was used and the samples were normalised to a next-to-next-to-leading-order (NNLO) prediction~\cite{Anastasiou:2003ds}.
 
The production of
\ttV\
events was modelled using the \MGNLO[2.3.3] generator at NLO with the \NNPDF[3.0nlo] PDF.
The events were interfaced to \PYTHIA[8.210]~ using the A14 tune and the \NNPDF[2.3lo] PDF set.
The production of tri-boson
($VVV$)
events was simulated with the \SHERPA[2.2.2] generator.
Matrix elements accurate to LO in QCD for up to one additional parton emission were matched and merged with the \SHERPA parton shower based on Catani--Seymour dipole factorisation using the \MEPSatNLO prescription.
Samples were generated using the \NNPDF[3.0nnlo] PDF set, along with the dedicated set of tuned parton-shower parameters developed by the \SHERPA authors.
 
The generated background MC samples were processed by the full ATLAS detector simulation 
based on \textsc{Geant4}~\cite{Agostinelli:2002hh}. 
The effect of multiple interactions in the same and neighbouring bunch crossings (\pileup) was modelled by overlaying the simulated hard-scattering event with inelastic $pp$ events generated with \PYTHIA[8.186]~\cite{Sjostrand:2007gs} using the \NNPDF[2.3lo] set of
PDF and the A3 set of tuned parameters~\cite{ATL-PHYS-PUB-2016-017}.
Simulated events were reweighted to match the pile-up conditions in the data.
All simulated events were processed using the same reconstruction algorithms and triggering requirements as used in data.


\section{Event Reconstruction and Pre-selection}
\label{sec:simu}

 
Proton-proton interaction vertices are reconstructed in events with at least two tracks, each with $\pT > 0.5~\GeV$.
The primary hard-scatter vertex is defined as the one with the largest value of the sum of squared track transverse momenta.
 
Muons are identified by matching tracks reconstructed in the MS to tracks reconstructed in the ID (referred to as combined muons).
To increase the muon reconstruction efficiency, non-combined muon identification algorithms are also used in the analysis,
including using the MS stand-alone tracks in the region $2.5 <|\eta|<2.7$,
and matching the ID tracks with calorimeter hit information within $|\eta| < $0.1,
as well as using the ID tracks associated with at least one local track segment in the MS.
In the $4\mu$ event selection at most one of the selected muons can be a non-combined muon.
Each muon is then required to satisfy the ‘loose’ identification criteria \cite{MUON-2018-03-updated}.
Muons are required to be isolated using a particle-flow algorithm \cite{PERF-2015-09}
and associated with the primary hard-scatter vertex by satisfying
$|{d_0 \over \sigma_{d_0} }| < $ 3
and $|z_0\times \sin\theta| <$ 0.5~mm,
where $d_0$ is the transverse impact parameter
calculated with respect to the measured beam-line position, $\sigma_{d_0}$ its uncertainty, and $z_0$ is the longitudinal distance between the point at which $d_0$ is measured and the primary vertex.
The minimum muon \pT threshold is 3 \GeV.
 
In addition to muons, electrons, jets and missing transverse momentum (\MET) are also used to select control samples for background estimation in this analysis.
The reconstructions of these objects are described below.
 
Electrons are identified with a likelihood discriminator built from the  shower shapes of electron energy deposits in the calorimeter, track-cluster matching, and some of the track quality distributions.
Each electron is required to satisfy the ‘medium’ likelihood identification criteria~\cite{Aad_2019}, as well as similar vertex and isolation requirements as muons.
The electrons are reconstructed in the region $|\eta| < $ 2.5, excluding the transition area between the barrel and endcap calorimeters, $1.37 < |\eta| < 1.52$, and required to have $p_{\textup{T}} >$ 7~GeV.
 
Jets are reconstructed with the anti-$k_t$ algorithm \cite{ATL-PHYS-PUB-2018-012, Fastjet} with a radius parameter of $R = 0.4$.
The jet clustering input objects are based on particle-flow \cite{PERF-2015-09} in the ID and the calorimeter.
Jets are required to have $p_{\textup{T}} > 20 ~\GeV$ and $|\eta|<2.5$.
Jets containing B hadrons, referred to as \bjets, are identified with a multivariate discriminant~\cite{ATL-PHYS-PUB-2016-012}.
To reduce the effect of pile-up, an additional quality requirement based on the `Jet-Vertex-Tagger' algorithm (JVT) \cite{ATLAS-CONF-2014-018} is applied in jet identification.
 
\MET is determined as the magnitude of the negative vectorial sum of the transverse momenta of the selected and calibrated physics objects (including muons, electrons, photons, and jets including hadronically decaying tau-leptons) and the ID tracks coming from the main vertex and not associated with any physics object (soft term)~\cite{PERF-2016-07}.
 
Events containing at least four muons with kinematics consistent with $Z(Z^*/\gamma^*)\rightarrow 4\mu$ production are then selected as follows.
The four leading \pT-ordered muons are required to pass the \pT thresholds of 20, 15, 8, and 3~GeV, respectively.
If a muon is selected as a non-combined muon, its \pT must be greater than 15 GeV.
Any di-muon pair in the event must have an invariant mass $m_{\mu\mu}$ greater than 4 GeV and an angular separation $\Delta R$ larger than 0.2.
To search for the $Z'\rightarrow \mu^+\mu^-$ signature, two muon pairs are selected based on their invariant mass values.
The first pair (referred to as $Z_1$) is selected from all the possible $\mu^+\mu^-$ pairs to have the smallest mass difference between the $Z_1$ mass and the $Z$ mass, $|m_Z - m_{Z_1}|$.
The second $\mu^+\mu^-$ pair is selected from the remaining muons that has the highest invariant mass (referred to as $Z_2$).
The correct signal di-muon pairing fraction varies with the \Zp mass, where the selected di-muon pair that forms $m_{Z1}$ or $m_{Z2}$ originates from the \Zp.
For example, for $m_{Z'}$ = 5, 42, 63, 72, and 81 GeV, the correct di-muon pairing fractions are about 78\%, 50\%, 88\%, 82\%, and 90\%, respectively.
Finally, the selected four muons must have an invariant mass in a range of 80 to 180 GeV, but excluding the Higgs boson resonance mass region of 110 to 130 GeV.
Including the 4$\mu$ events in the 4$\mu$ mass region between 130 to \SI{180}{\GeV} improves the $Z'$ signal detection efficiency for its mass greater than 60 \GeV.

The $Z'$ signal efficiency at various stages of the event selection is shown in Table~\ref{tab:cutgflow} for five representative mass points.
The event selection efficiencies vary significantly depending on the $Z'$ mass.
At generator level,
an MC filter is applied, which requires at least four muons with $p_T > 2$~GeV and $|\eta| < $ 3.0.
The MC filter efficiencies of these representative $Z'$ signal samples are listed in the table as well.
 
\begin{table}[!h]
\caption{
The $Z'$ signal event selection efficiencies compared to the events passing the previous cut level for several representative mass points.
The overall signal efficiencies are the products of the 4$\mu$ MC filter and the combined
event selection efficiencies. 
}
\centering
\begin{tabular}{l|ccccc}
\toprule
$m_{Z'}$ [GeV]                                    & 5      & 42      & 63     & 72     & 81     \\
\midrule
\midrule
MC filter efficiency                              & 32.8\% & 57.7\%  & 61.0\% & 65.3\% & 70.0\% \\
\midrule
Number of identified muons $\geq$ 4               & 47.3\% & 74.1\%  & 70.8\% & 72.4\% & 75.4\% \\
$p_{\textup{T}}^i (i=1,4) > 20, ~15, ~8, ~3$~GeV  & 60.0\% & 82.6\%, & 90.3\% & 93.6\% & 98.2\% \\
$\Delta R(\mu_i,\mu_j)>0.2$ \& vertex requirement & 87.2\% & 95.4\%  & 96.2\% & 96.6\% & 97.2\% \\
Isolation                                         & 54.2\% & 76.9\%  & 79.2\% & 84.1\% & 87.5\% \\
$m_{4\mu}$ within [80, 110] or [130, 180] GeV     & 91.9\% & 88.8\%  & 58.9\% & 33.5\% & 16.8\% \\
\midrule
Combined event selection efficiency               & 12.3\% & 39.9\%  & 28.7\% & 18.4\% & 10.6\% \\
 
\midrule
Overall 4$\mu$ signal efficiency                  & 4.1\%  & 23.0\%  & 17.5\% & 11.9\% & 7.4\%  \\
\bottomrule
\end{tabular}
\label{tab:cutgflow}
\end{table}
 
The $Z'$ production signature is searched for in the $Z_1$ or $Z_2$ mass spectrum depending on the assumed $Z'$ mass.
The relatively high-mass $Z'$ signals mostly appear as a peak in the $Z_1$ spectrum while the relatively low-mass signals mostly appear as a peak in the $Z_2$ spectrum.
Representative examples of the predicted signal over background, after further selection with a deep learning approach which will be described in \Sect{\ref{sec:dnn}},
are shown in \Fig{\ref{fig:bkg-only-Z1-Z2}}.
In the analysis the $Z_1$ and $Z_2$ mass spectra are scanned to search for a \Zp with mass greater or smaller than 42~\GeV, respectively.
The boundary value of 42~GeV is chosen based on the studies to optimize the search sensitivity.

The numbers of 4$\mu$ events in data and the estimated background yields are given in \Tab{\ref{tab:background-estimate}}.
More details about  the estimation of the reducible backgrounds containing non-prompt muons can be found in \Sect{\ref{sec:background}}.
The total uncertainties of simulated backgrounds are also listed in the table.
The evaluations of systematic uncertainties will be described in \Sect{\ref{sec:syst}}.
 
\begin{table}[htb]
\caption{
The selected 4$\mu$ events in data and the estimated backgrounds and their combined statistical and systematic uncertainties.
}
\centering{
\begin{tabularx}{\textwidth}{c|c|cccc}
\toprule
Data               & Total      & $qq\rightarrow ZZ^*$ & $gg\rightarrow ZZ^*$ & $ttV+VVV+H$ & Reducible background \\
& background & from MC              & from MC              & from MC     & from data            \\
\midrule
1131               &
$1148 \pm 70$      &
$1065^{+70}_{-69}$ &
$15.6 \pm 2.5$     &
$6.2 \pm 2.9$      &
$61.1 ^{+8.3}_{-9.1}$                                                                                              \\
\bottomrule
\end{tabularx}
}
\label{tab:background-estimate}
\end{table}


\section{Estimation of Reducible Background from Data }
\label{sec:background}


A data-driven technique is used to estimate the reducible background from \Zjet and \ttbar production.
Events from these processes may contain two prompt leptons from $W$ or $Z$ boson decays, together with additional activity such as heavy-flavor jets or misidentified components of jets yielding reconstructed muons, collectively referred to as non-prompt muons.
Such backgrounds are estimated from data using a control sample of $\mu^+\mu^- \mu_j \mu_j$ events, selected with the standard signal requirements except that non-prompt muons, $\mu_j$, are selected in place of two of the signal muons. Non-prompt muons are defined as muon candidates using the standard selection but fail the requirements on isolation and impact parameter. However, they are required to pass a much looser impact parameter significance requirement, $|{d_0 \over \sigma_{d_0}} |< 10$.
The background in the signal sample is estimated by scaling each event in the selected $\mu^+\mu^- \mu_j \mu_j$ control sample by $f_1 \times f_2$, where the $f_i (i=1, 2)$ is referred to as a "fake-factor" for each of the two non-prompt muons and is computed as a function of their $p_{\textup{T}}$ and $|\eta|$.
 
The fake-factor $f$ is derived from two independent non-prompt muon enriched data control samples dominated by \Zjet or \ttbar events.
The \Zjet (\ttbar) sample contains non-prompt muons dominantly from light jets (heavy-flavor jets).
The \Zjet sample is tagged by two isolated prompt leptons from \Zboson decays ($Z\rightarrow \mu^+\mu^-, ~ e^+e^-$).
The \pT-threshold for the two prompt leptons are 27 and 25 GeV and their invariant mass must be within 10 GeV of the \Zboson mass.
The \MET of the event must be less than 25 GeV.
The \ttbar sample is tagged by two prompt isolated high- \pT $e^\pm \mu^\mp$ leptons with \pT thresholds of 28 and 25 GeV and associated with at least one $b$-jet.
The event is required to have $\MET >$ 50 GeV.
If there is an additional isolated high-\pT lepton in the event, the transverse mass calculated with the \MET and the third lepton is required to be less than 60 GeV to reduce the contribution from the $t\bar t W$ events.
 
Events with four muons that satisfy the signal selection criteria are removed from both \Zjet and \ttbar control samples.
In addition, MC simulated diboson events satisfying the selection criteria are subtracted from these control samples.
Excluding the target leptons,
the additional reconstructed muon objects, passing either the signal muon or the non-prompt muon selection criteria are counted in the control samples.
The fake-factor $f$ is calculated as the ratio of the
probability for a non-prompt muon to satisfy the signal muon selection criteria to the probability
for a non-prompt muon to fail the signal muon selection.
The factor $f$ is derived as a function of $|\eta|$ and \pT of the non-prompt muons,
which varies from 0.11 to 0.76 (0.01 to 0.27) in the \Zjet (\ttbar) control sample.
Then, the fake-factors derived from two data control samples are combined using the $m_{Z1}$ spectra obtained from simulated $4\mu$ background events from \Zjet and \ttbar processes.
A simultaneous fit of the $m_{Z1}$ spectra of the MC events to data selected in the $\mu^+\mu^- \mu_j \mu_j$ control sample is performed to determine the fractions of the \Zjet and \ttbar events in each bin of the mass spectrum.
 
The overall systematic uncertainty of $f$ is about 14\%.
It is determined with alternative non-prompt muon selections, such as changing the muon isolation criteria in data control samples (7.6\%), and the uncertainties of the \Zjet and \ttbar event fractions when combining the fake-factors derived from two control samples (8.6\%).
The statistical uncertainties of the control samples are also accounted as part of the systematic uncertainties (6.1\%).
 
Additional reducible background contributions come from the $WZ$ process. These events contain three prompt muons from $W$ and $Z$ decays and one non-prompt muon. This background is estimated by scaling a simulated $3\mu +\mu_j$ sample with the $f$ derived from the \Zjet sample.
 
The total estimated number of reducible background events is
$61.1^{+8.3}_{-9.1}$   
(with 88\% and 12\% contributions from the $\mu^+\mu^- \mu_j \mu_j$ and $ 3\mu +\mu_j$ control samples, respectively)
as listed in Table~\ref{tab:background-estimate}.


\section{Event Classification with Deep Learning Approach}
\label{sec:dnn}

The selected 4$\mu$ events are classified with a "Deep Learning" approach to further separate the $Z'$ signal from the SM background.
A parametrized deep neural network (pDNN) architecture~\cite{Baldi_2016} is used in the analysis.
This algorithm allows the training of a single classifier for multiple signal mass hypotheses in the search range.
In practice, the kinematic inputs together with the $Z'$ mass parameters of signal and background are used for training.
The MC generated $Z'$  masses (listed in Table~\ref{tab:signalSamples}) are used as the multiple signal mass parameters,
while the distribution of the mass parameter for background events is randomly drawn from the same distribution as for the signal class.
The algorithm was implemented in the Keras \cite{chollet2015keras} framework with the TensorFlow \cite{tensorflow2015-whitepaper} backend.
Two classifiers are trained for low (high) $Z'$ mass searches using MC mass parameters lower (higher) than 40 GeV.
During the training of the classifier, the training samples were composed of simulated signal and background 4$\mu$ events using the pre-selection described in Section~\ref{sec:simu}.
A set of kinematic distributions were used for pDNN training input features.
They are the \pT and $\eta$ of each muon, the \pT of the $Z_1$ and $Z_2$, the mass difference of the $Z_1$ and $Z_2$, $\Delta R$ and $\Delta\eta$ of each muon pair that forms the $Z_1$ and $Z_2$, and the \pT and mass of the $4\mu$ system.
Examples of the input variable distributions, both for representative signals ($m_{Z'}$ = 15, and 51 GeV) and background, are shown in \Fig{\ref{fig:Z1Z2}} to compare the predicted and observed
transverse momenta ($p_{\textup{T}}^{Z_1}$ and $p_{\textup{T}}^{Z_2}$) distributions
and the mass difference of the $Z_1$ and $Z_2$.
In addition to the major background from the SM $Z(Z^*)\rightarrow 4\mu$ production, other backgrounds, including 4$\mu$ events containing non-prompt muons estimated from data, and from $ttV$, $VVV$, and Higgs boson production processes, are included in the plots.
 
\begin{figure}[!htbp]
\centering
 
\subfloat[]{
\includegraphics[width=0.33\textwidth]{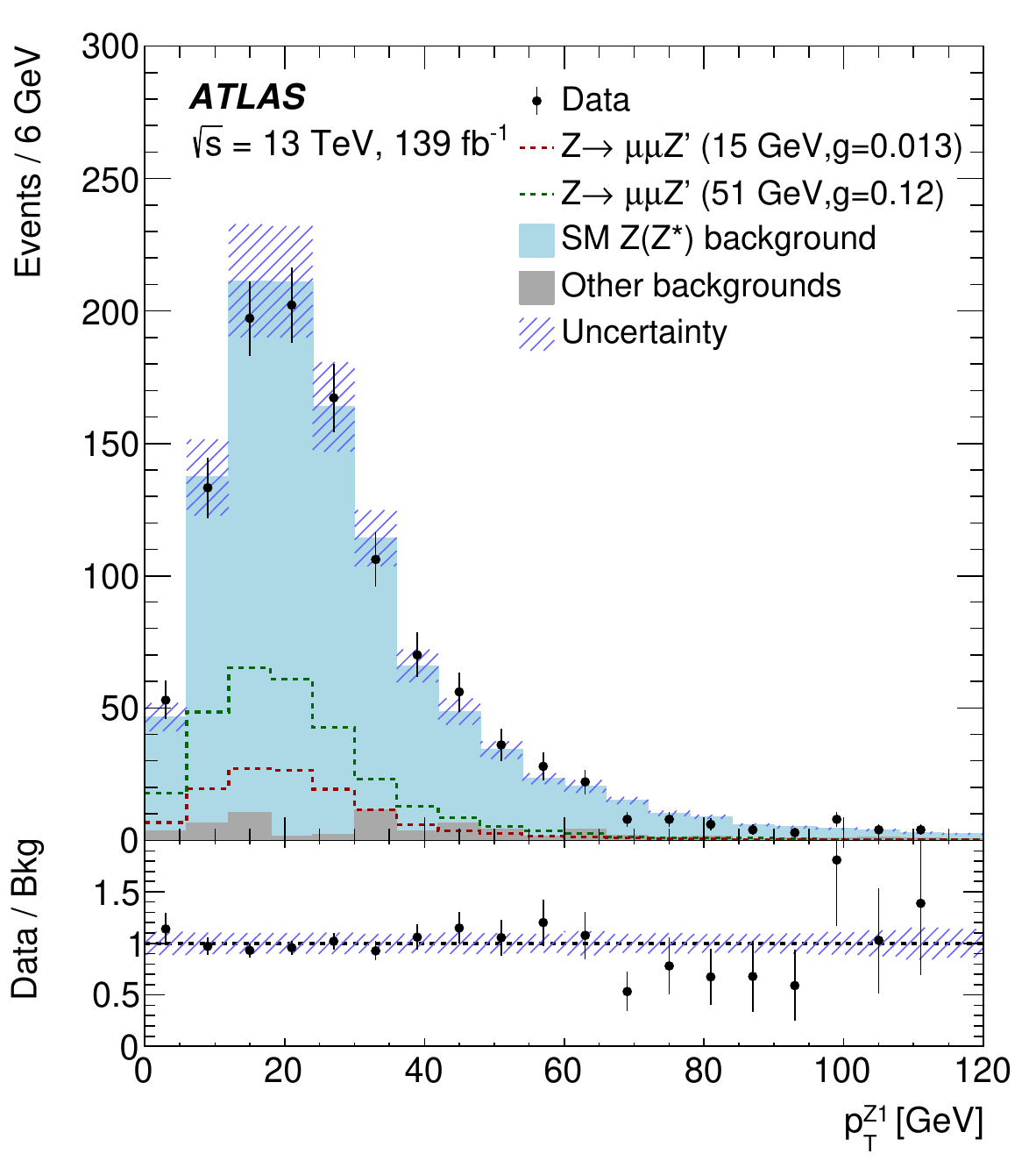}
}
\subfloat[]{
\includegraphics[width=0.33\textwidth]{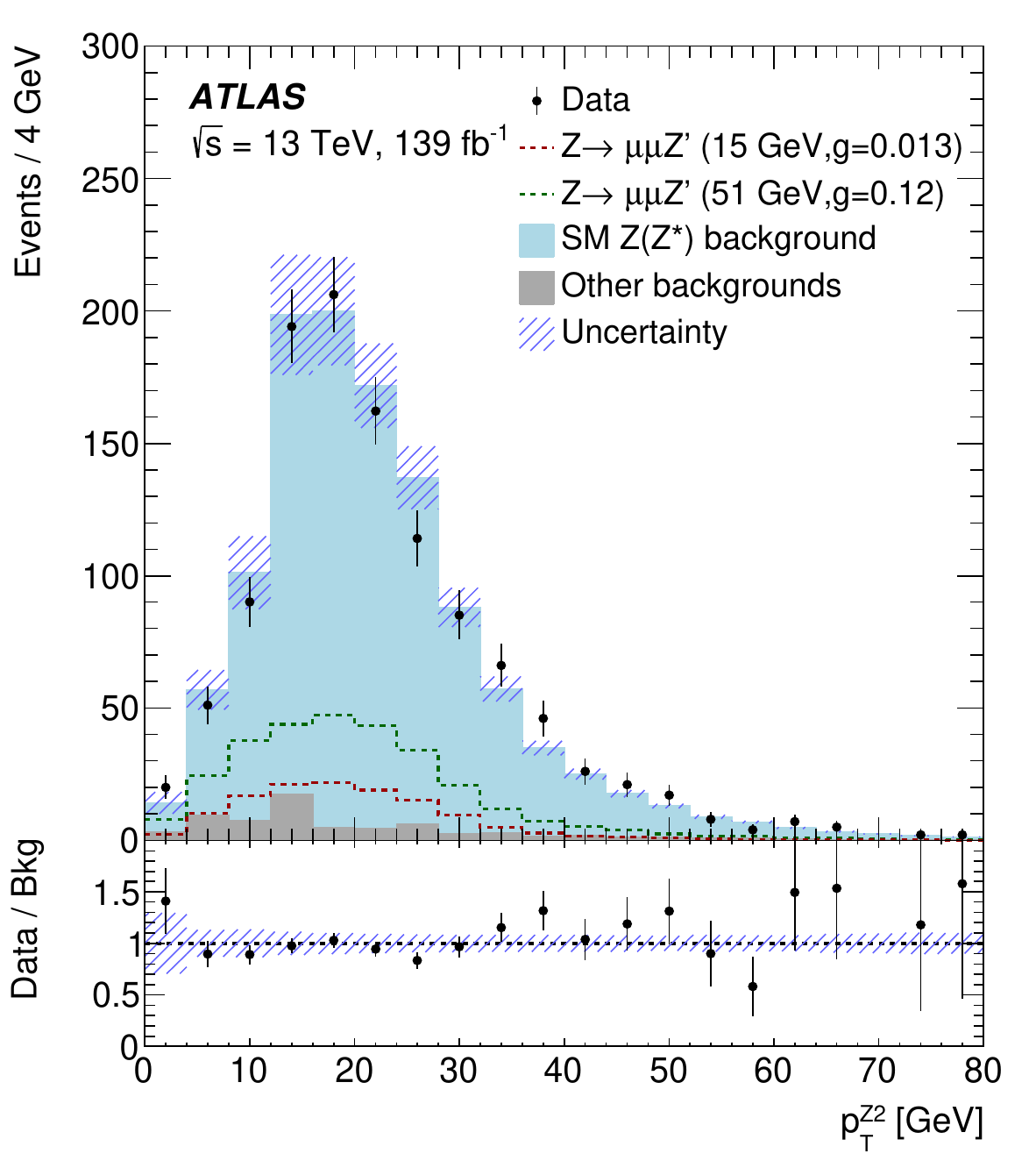}
}
\subfloat[]{
\includegraphics[width=0.33\textwidth]{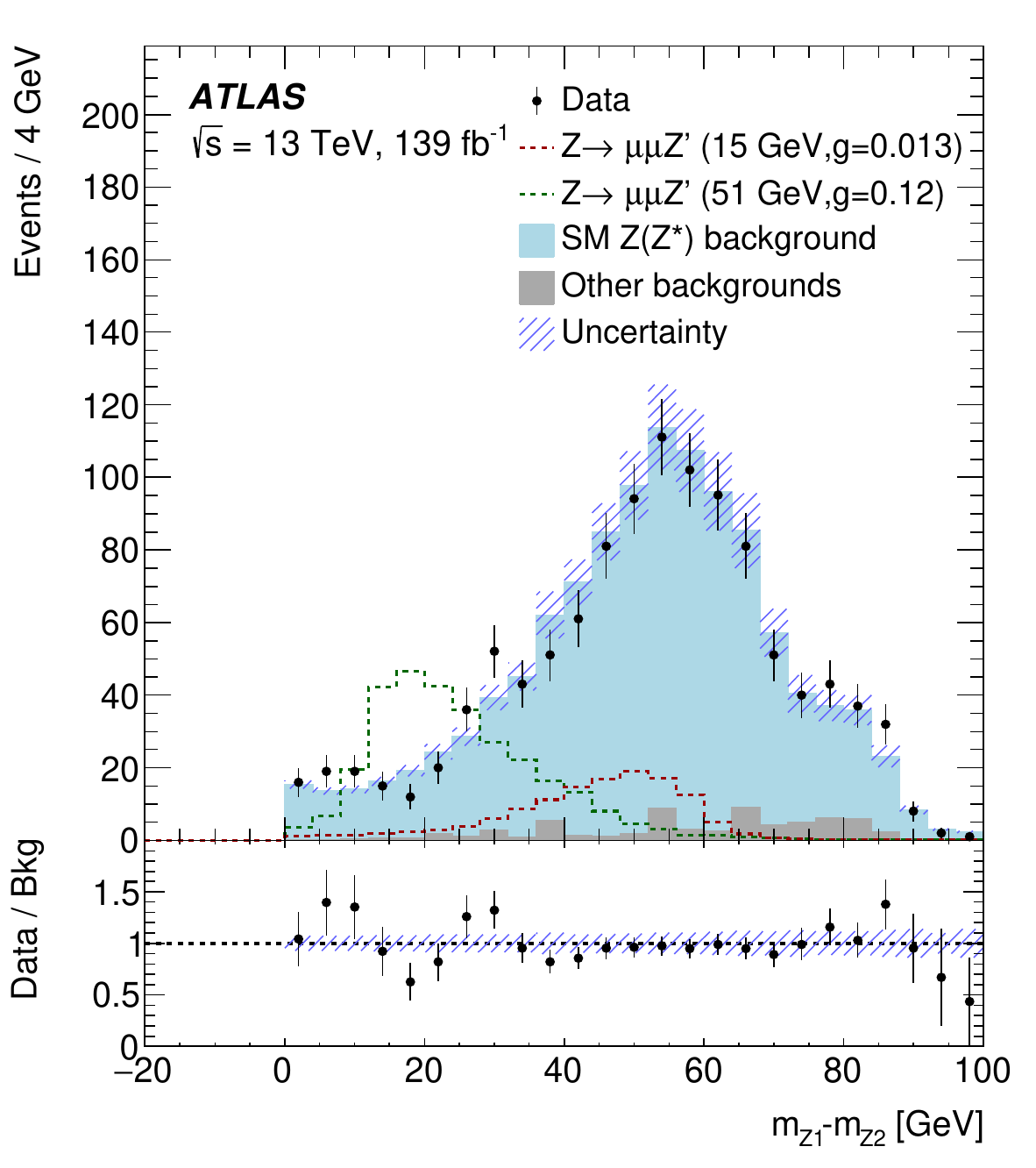}
}
\caption{Distributions of $p_{\text{T}}^{Z_1}$ (a), and $p_{\text{T}}^{Z_2}$ (b), and
the mass difference of the $Z_1$ and $Z_2$ candidates (c). 
Small background contributions are denoted as "other backgrounds", including 4$\mu$ events containing non-prompt muons estimated from data and from $ttV$, $VVV$, and Higgs boson production processes.
}
\label{fig:Z1Z2}
\end{figure}
 
The training fits the weight parameters of the pDNN models which are then applied to real data or MC samples to obtain the pDNN output discriminating variable.
The $m_{Z1}$ or $m_{Z2}$ are used as the mass parameter when applying the model to data.
 
\begin{figure}[!hb]
\centering
\subfloat[low mass: $m_{Z'} \leq 42~\GeV$]{
\includegraphics[width=0.48\textwidth]{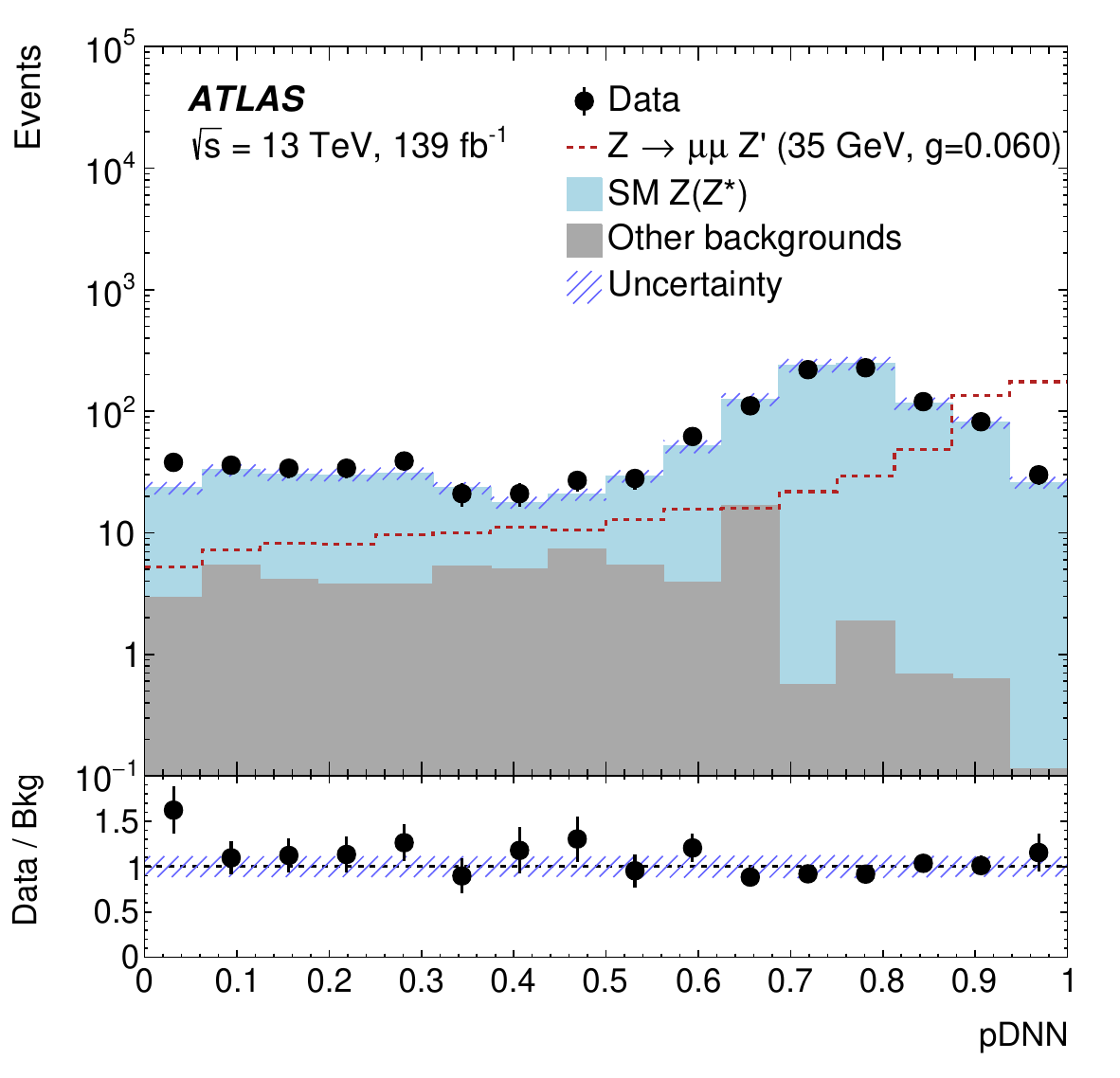}
\label{fig:pnn_score:low}
}
\subfloat[high mass: $m_{Z'} \geq 42~\GeV$]{
\includegraphics[width=0.48\textwidth]{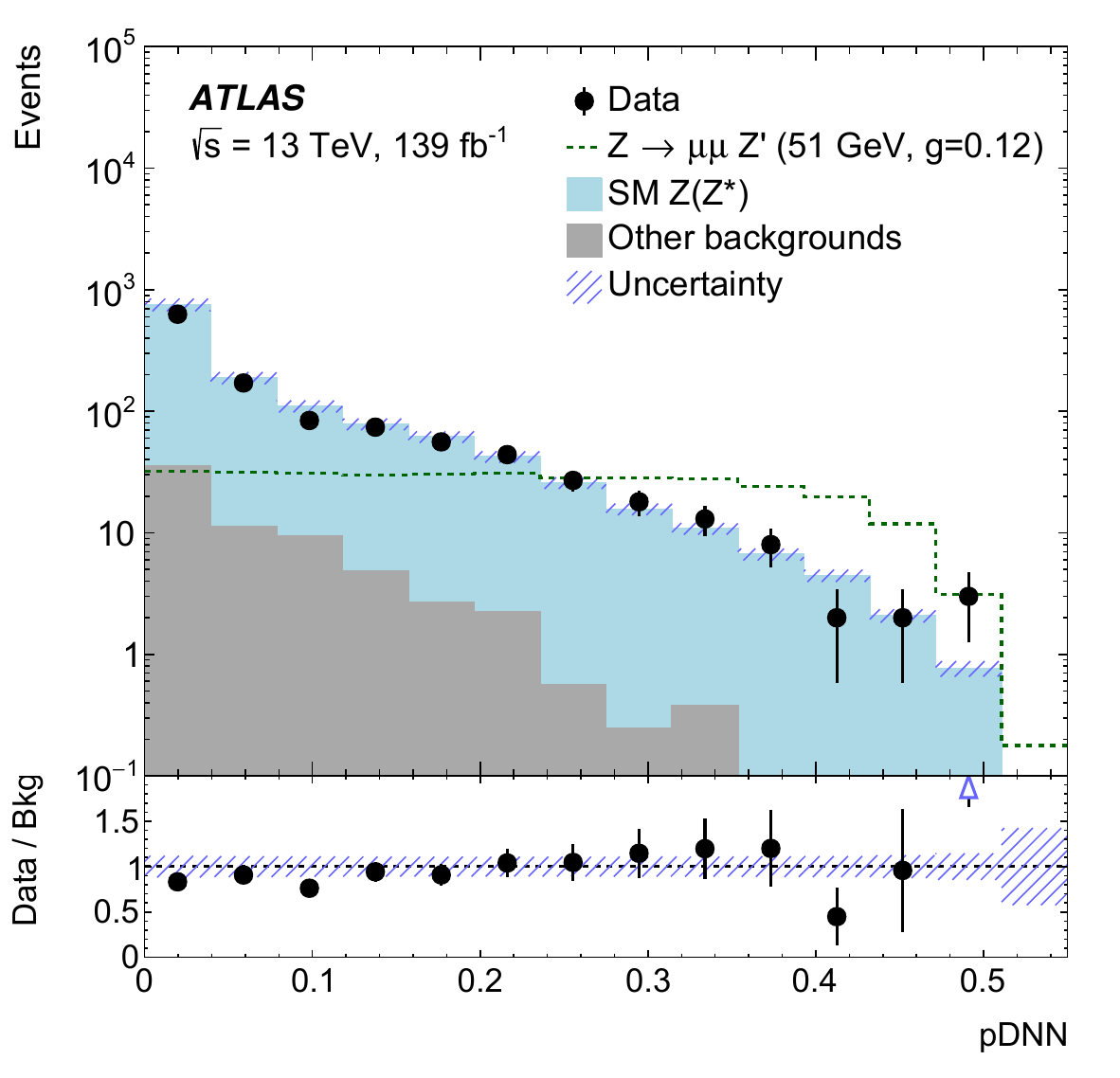}
\label{fig:pnn_score:high}
}
\caption{
The pDNN output discriminant variable distributions for low mass (a) and high mass (b) with a signal sample at 35 \GeV ~and 51 \GeV, respectively.
}
\label{fig:pnn_score}
\end{figure}

To improve the power of the pDNN classifier, the model is optimized through an automatic process developed with the package \textsc{Tune}~\cite{liaw2018tune} and \textsc{HEBO Search} algorithm~\cite{alex2020hebo}.
The neural-network structure was chosen to have two fully connected hidden layers, each with 256 (32) nodes for the high (low) mass search in the analysis.
Other training hyper-parameters, such as the \textit{learning rate} with \textit{decay} and \textit{class weight}, for both high and low mass searches were also selected.
Two pDNN output discriminant classifiers for high
and low mass
searches were obtained and are shown in Figure~\ref{fig:pnn_score} for data and simulated signal and background events.

To maximize the search sensitivity, a scan of the pDNN output scores was performed to find the optimal cut values for each \Zp mass hypothesis.
These cut values vary from 0.42 to 0.74 (0.12 to 0.16) for the low (high) mass region.
These cuts keep high signal efficiencies between 98\% and 95\%  (90\% and 50\%), while the corresponding background reductions range from 10\% to 50\%  (50\% to 96\%) for the low (high) mass region.

After the 4$\mu$ event selection with the pDNN classifier, the final discriminant to search for the $Z'$ resonance signature is the $Z_1$ (for $m_{Z'} \geq 42$~GeV) or $Z_2$ (for $m_{Z'} \leq 42$~GeV) mass spectrum as shown in \Fig{\ref{fig:bkg-only-Z1-Z2}}.
Data are compared to the estimated background together with two representative signals with masses of 15 and 51 GeV, shown in \Fig{\labelcref{fig:bkg-only-Z1-Z2:a,fig:bkg-only-Z1-Z2:b}}, respectively.
The values of the gauge coupling strengths ($g$) for the two mass points are chosen for the purpose of illustration.
 
One should note that the dimuon masses ($m_{Z1},~ m_{Z2}$) are not used as the pDNN training variables, therefore cutting on the pDNN output scores does not sculpt the dimuon invariant mass distributions of the backgrounds (see \Fig{\ref{fig:bkg-only-Z1-Z2}}).
To interpret the observed mass spectra, systematic uncertainties in the predictions are estimated in Section~\ref{sec:syst}. 
 
\begin{figure}[!htb]
\centering
\subfloat[]{
\includegraphics[width=0.48\textwidth]{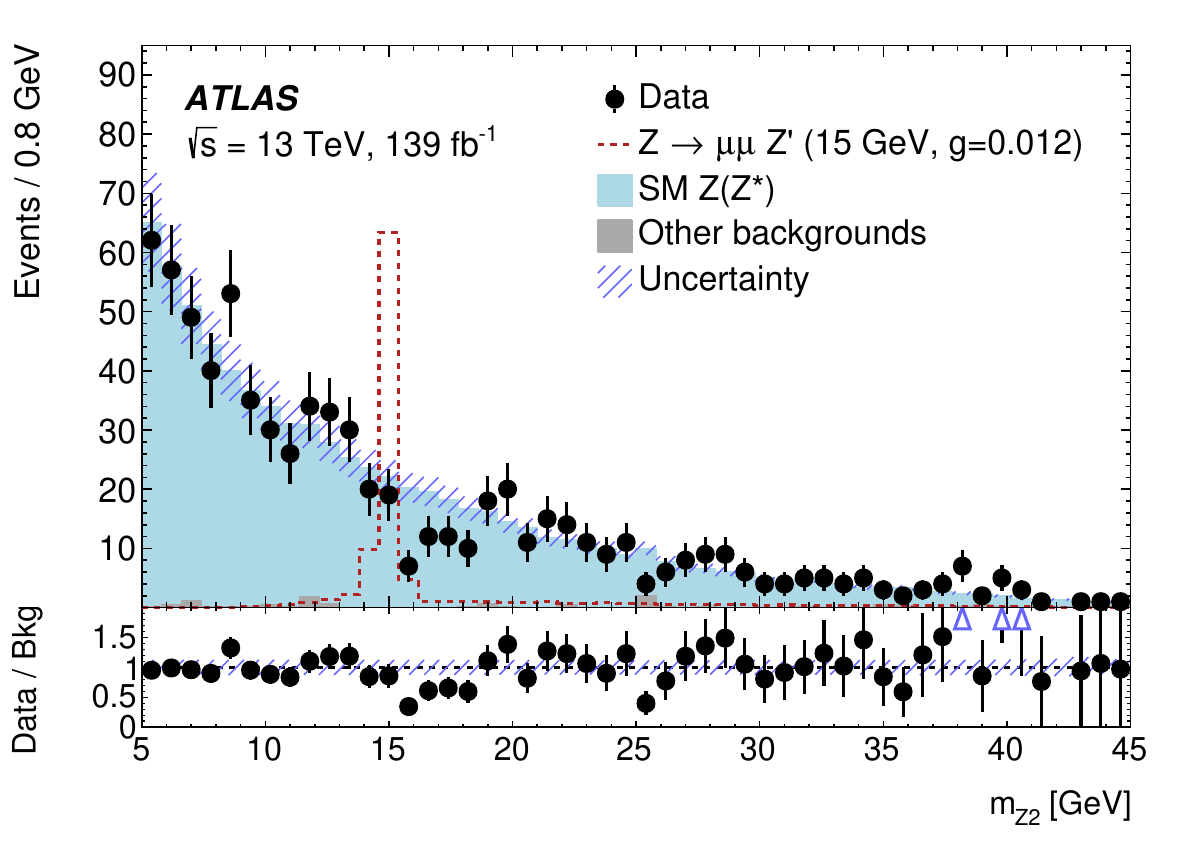}
\label{fig:bkg-only-Z1-Z2:a}
}
\subfloat[]{
\includegraphics[width=0.48\textwidth]{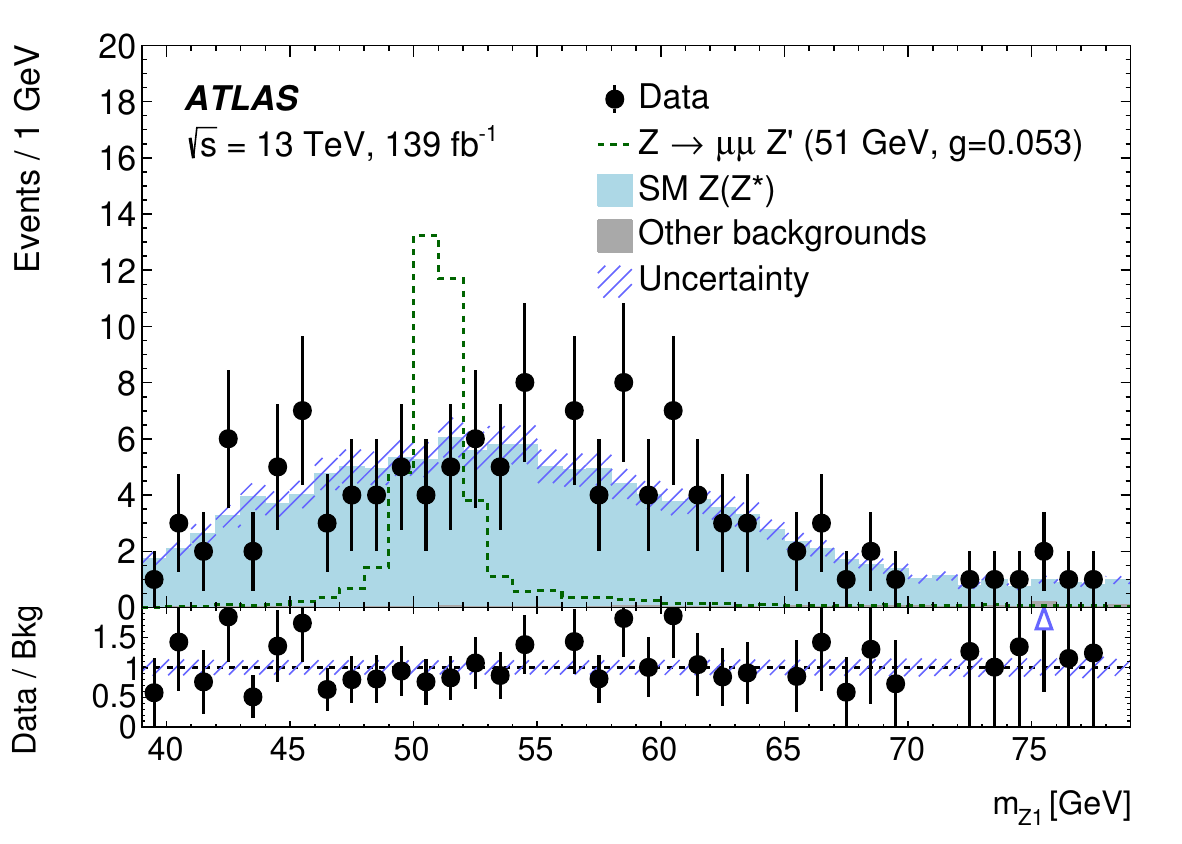}
\label{fig:bkg-only-Z1-Z2:b}
}
\caption{
Mass spectra of $m_{Z2}$ (left) and $m_{Z1}$ (right) for the pDNN-selected events with a signal sample at 15 \GeV ~and 51 \GeV, respectively.
}
\label{fig:bkg-only-Z1-Z2}
\end{figure}


\section{Systematic Uncertainties}
\label{sec:syst}

 
Systematic uncertainties in the simulated event yields and shapes, for both signal and background processes, may arise from the calibration of the physics objects and from uncertainties in theoretical modelling used in the predictions.
 
The major experimental uncertainties come from the muon reconstruction, identification, and isolation requirement efficiencies.
These efficiencies are corrected based on studies performed in data control regions.
The energy and momentum scales and resolutions of the simulated objects are
corrected to reproduce data from $Z\rightarrow \mu^+\mu^-$ and $J/\psi \rightarrow \mu^+\mu^-$ decays~\cite{MUON-2018-03-updated}.
The uncertainties on the SM 4$\mu$ detection efficiency are determined
by varying the nominal calibrations in the MC samples by one standard deviation, including muon momentum resolutions and scales, and the trigger, reconstruction, identification, and isolation requirement efficiencies.
The overall relative experimental uncertainties in the SM 4$\mu$ background event selection efficiency is about 3.9\%, dominated by the uncertainty of the isolation efficiency (2.9\%) and the low $p_{\textup{T}}$ calibration uncertainty (2.0\%).
The hypothetic \Zp signal event selection efficiency uncertainties vary from 8.3\% to 3.9\%, depending on the \Zp mass.
In addition, the uncertainty in the combined 2015-2018 integrated luminosity is 1.7\%~\cite{ATLAS-CONF-2019-021}, obtained using the LUCID-2 detector~\cite{Avoni_2018} for the primary luminosity measurements.
 
Sources of theoretical uncertainties come from the choice of QCD scales (renormalization $\mu_R$ and factorization $\mu_F$), strong coupling constant $\alpha_S$, and PDFs, as well as the parton shower models.
These uncertainties affect the signal and background event selection efficiencies, normalization, and the shape of their kinematic distributions.
The scales are varied independently from 0.5 to 2.0 times the nominal values and the largest deviation is chosen as the systematic uncertainty.
The PDF uncertainty is estimated by comparing events generated with different PDF sets, as well as the uncertainties from the nominal PDF set itself.
The maximal variation (envelope) is accounted for as the systematic uncertainty, following the PDF4LHC~\cite{Butterworth:2015oua} recommendations.
The $\alpha_{S}$ uncertainty is estimated by comparing events generated with different $\alpha_{S}$ values using the nominal PDF set.
The parton shower uncertainty is estimated by comparing events with different parton shower parameters in the Sherpa MC samples.
For the $Z(Z^*)\rightarrow 4\mu$ process, the relative uncertainties of event yields for scale, PDF, $\alpha_s$, and parton shower,  are 4.6\%, ~1.8\%, ~1.0\%, and 1.9\%, respectively.
The uncertainties of small background contributions from
the $t\bar t V$ ($5.2 \pm 0.7$),
$VVV$ ($0.5 \pm 0.25$),
and Higgs boson ($0.53 \pm 0.06$)
processes are evaluated and  included in the analysis.
 
The interference effect between the signal and background is estimated (using generator-level MC samples simulated with \MGNLO)
by evaluating the cross-section ratio, ${\Delta \sigma} / {\sigma_{Z'}}$,
where $\Delta \sigma$ is
the difference of the cross-section from the inclusive MC sample and the sum of the cross-sections of the signal and background samples in the $Z'$ detection phase space.
The effect varies from 1.8 to 7.5\% for the $Z'$ mass from 5 - 81 GeV,
which is accounted for as additional signal yield systematic uncertainties.
In addition, the MC $Z'$ signal filter acceptance uncertainty is estimated to be about 2\%, which
is calculated by varying the QCD scales, the PDF-sets, and the strong coupling constant using MC events at the generator-level for different $Z'$ mass points.

All the uncertainties (from both experiment and theory) on the final discriminant di-muon mass spectra are included as nuisance parameters in the signal extraction fitting process described in Section~\ref{subsec:final_result}.


\section{Data Interpretation and Results}
\label{sec:result}


The statistical analysis is performed by comparing the data to
the sum of the background prediction and the signal
to search for the $Z'$ signature and information about the $pp\rightarrow \mu^+\mu^-Z'\rightarrow 4\mu$ signal production cross-section and the associated coupling strength.
In case of no significant data excess over the background prediction, upper limits on the signal production cross-section times the decay branching fraction for different \Zp masses are set at 95\% confidence level (CL).

\subsection{Statistical Procedure and Results}
\label{subsec:final_result}
 
The statistical procedure used to interpret the data is described in \Refn{\cite{HIGG-2012-17}}.
The likelihood function is constructed from the product of Poisson probabilities:
\begin{equation*}
L (\text{data} |~\mu, \theta)= \prod_{i=1}^N \text{Poisson}(\text{data}_i |~\mu \cdot s_i(\theta) + b_i(\theta))\times {\text G}(\tilde\theta |\theta)
\end{equation*}
where $i$ is the bin-index of the fitted variable distribution;
$\mu$ is the signal strength, and $\theta$ denotes the nuisance parameters, which represent the uncertainties of the measurements;
${\text G}(\tilde\theta |\theta)$ is the Gaussian scaling function
of the nuisance parameters constructed as deviations from the nominal model of the systematic uncertainties, where $\tilde\theta$ provides a maximum likelihood estimate for $\theta$.
The parameter of interest in the statistical analysis is the global signal strength factor $\mu$, which acts as a scale factor on the total number of events predicted by the signal model.
This factor is defined such that $\mu = 0$ corresponds to the background-only hypothesis and $\mu > 0 $ corresponds to a \Zp signal in addition to the background.
Hypothesised values of $\mu$ are tested based on the profile likelihood ratio~\cite{Cowan:2010js}, which compares data with background-only ($b$) and signal+background ($s+b$) models using the following test statistic:
\begin{equation*}
q_\mu =
\begin{cases}
-2 \text{ln} \frac{L( \text{data} |~\mu, \hat{\hat{\theta}}_\mu)}{L(\text{data} |~\hat\mu, \hat\theta)} & \hat{\mu} \leq \mu \\
0                                                                                                       & \hat{\mu} > \mu    \\
\end{cases}
\end{equation*}
where $\hat\mu ~\text{and} ~\hat\theta$ are the values
for $\mu$ and $\theta$ when maximizing the $L$ with all parameters floating, which are named as the unconditional maximum-likelihood (ML) estimators;
$\hat{\hat{\theta}}_\mu$ is the conditional ML estimator of $\theta$ for a fixed value of $\mu$.
This test statistic extracts the information on the signal strength from a full likelihood fit to the data.
The likelihood function includes all the parameters that describe the systematic uncertainties and their correlations.
 
The significance of an excess in the data is first quantified
with the local $p_0$, the probability that the background
can produce a fluctuation greater than or equal
to the excess observed in data.
The equivalent formulation in terms of number of standard deviations is
referred to as the local significance, which is computed from test statistic $q_0$:
\begin{equation*}
p_0 = \text{P}(q_0 < q_0^{obs}).
\end{equation*}
The global probability
for the most significant excess to be observed anywhere in a given search region is estimated with the method described in Ref.~\cite{Gross_2010}.
 
In case of no significant data excess over the background, exclusion limits are set, based on the CLs prescription~\cite{Read:2002hq}, which calculates the ratio of the probabilities based on the $s+b$ background-only models:
\begin{equation*}
\text{CL}_s(\mu) = \frac{\text{CL}_{s+b}}{\text{CL}_{b}}
= \frac{\text{P}(q_\mu> q_\mu^{obs})}{\text{P}(q_0 > q_0^{obs})}
\end{equation*}
A value of $\mu$ is regarded as excluded at 95\% confident level when CLs is less than 5\%.
 
The statistical tests are performed by first searching for data excesses by means of a $p_0$ scan, and in case no such significant data excess is found, by setting limits on the \Zp production cross-section times the branching fraction.
In both steps the values and reconstructed mass distributions of hypothesized \Zp mass $m_{Z'}$ are used.
The steps between the mass points used in these test procedures are set to match the mass resolutions determined by reconstruction of the di-muon invariant mass.
A linear interpolation of the expected event yields and shapes between the MC generated signal samples is used in the fitting process.
The asymptotic approximation~\cite{Cowan:2010js} upon which the results are based has been validated against the method (detailed in \Refn{\cite{HIGG-2012-17}}) of using pseudo-experiments for several mass points.
 
In both steps of the statistical tests, data are fit to the $m_{Z1}$ and $m_{Z2}$ spectrum with background ($b$) only and signal+background ($s+b$) hypotheses.
In the fitting process each mass spectrum is divided in the signal region (SR) and the background control region (CR).
For each $Z'$ mass point the SR is defined in a mass window of $m_{Z'} \pm 3\sigma_{m_{\mu\mu}}$ of the di-muon mass spectrum.
The sidebands outside of the SR are defined as the CR.
The di-muon mass resolution $\sigma_{m_{\mu\mu}}$ is determined by the fully simulated $Z'$ mass distribution, which combines the \Zp natural width and the detector resolution, ranging from 0.10 to 1.75 GeV.
The mass resolution is mostly dominated by the detector resolution.
Finer binning is used in the SR to enhance the sensitivity.
The background CR is used to constrain the overall normalization for the background in the signal region.
The shape of the major background from $Z(Z^*)\rightarrow 4\mu$ is fixed with prior uncertainties included in the fitting process, but the normalization (or strength) floats in the fit.
Other background normalizations and shapes are fixed with prior uncertainties included in the fitting.
 
\subsection{The $p_0$ Scan Results}
\label{subsec:p-scan}
 
The $p_0$-values corresponding to the background-only hypothesis are scanned in the mass range of this analysis.
A binned profile-likelihood fit \cite{Cowan:2010js} is performed simultaneously across the $Z'$ signal-region and the background control region using the predicted and observed mass spectrum as inputs.
Data are fit to the $m_{Z1}$ and $m_{Z2}$ distributions
for $m_{\Zp} \geq 42~\GeV$ and $m_{\Zp} \leq 42~\GeV{}$, respectively,
with a "sliding" mass window as the defined SR changes for different \Zp mass points.
The chosen bin-size inside the SR is around 0.3~$\sigma_{m_{\mu\mu}}$, for each mass point. The total number of bins in the CR is 20.
The fit mass range of $m_{Z1}$ ($m_{Z2}$) is [30, 85] GeV ([0, 45] GeV).
 
The $p_0$-values at different \Zp mass hypothesis points are
computed and transformed into Gaussian standard deviations to indicate the significance as
shown in \Fig{\ref{fig:p_value_scan}}.
The smallest $p_0$-value is at 39.6 GeV, corresponding to a local 2.65$\sigma$ deviation from the background-only hypothesis, while the global deviation~\cite{Gross_2010} is found to be 0.52$\sigma$, indicating that no significant data excess over the expected background is observed.
 
\begin{figure}[!htbp]
\centering
\includegraphics[width=0.6\textwidth]{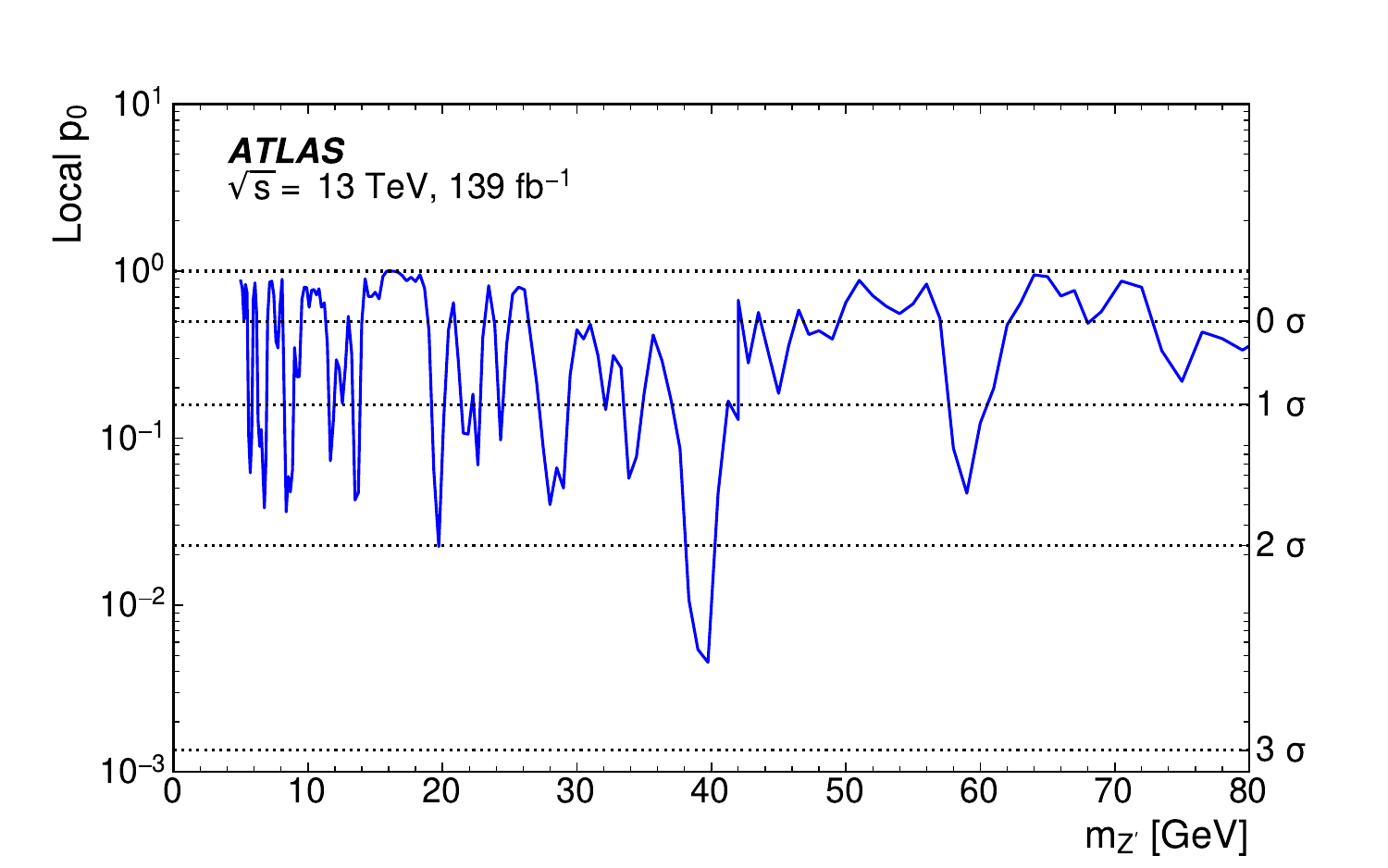}
\caption{The $p_0$-value scan across the \Zp mass signal regions.}
\label{fig:p_value_scan}
\end{figure}

\subsection{Upper Limits}
\label{subsec:limits}
 
The upper limits on the production cross-section times branching fraction of the $pp\rightarrow \mu^+\mu^-Z'\rightarrow 4\mu$ process are calculated using a similar fitting procedure described in Section~\ref{subsec:p-scan}.
Confidence intervals are computed based on the profile-likelihood-ratio test statistics~\cite{Cowan:2010js}.
The observed and expected upper limits at 95\% CL on the cross-section times branching fraction, $\sigma (pp\rightarrow Z'\mu\mu\rightarrow 4\mu)$, are shown in \Fig{\ref{fig:xs_limit}}.
The upper limits on the coupling parameter \coupzp are extracted from the limits of the $Z'$ production cross-section times branching fraction using the $L_\mu-L_\tau$ model,
which is determined by counting all the possible \Zp decay modes in this model.
At each generated $Z'$ mass point, a limit on the coupling strength \coupzp has been obtained from the cross-section limit.
The observed and expected upper limits on the coupling parameter \coupzp are shown in \Fig{\ref{fig:coup_limit}}.
The limits on the coupling \coupzp are in the range of 0.003 (for $m_{Z'}$ = 5~GeV) to 0.2 (for $m_{Z'}$ = 81 GeV) depending on the \Zp mass ranging from 5 to 81 GeV.
This ensures that the ratio of the $Z'$ natural width and mass, $\Gamma(Z') / m_{Z'}$, is well below 1\% in this mass range.
\begin{figure}[!htbp]
\centering
\subfloat[]{
\includegraphics[width=0.49\textwidth]{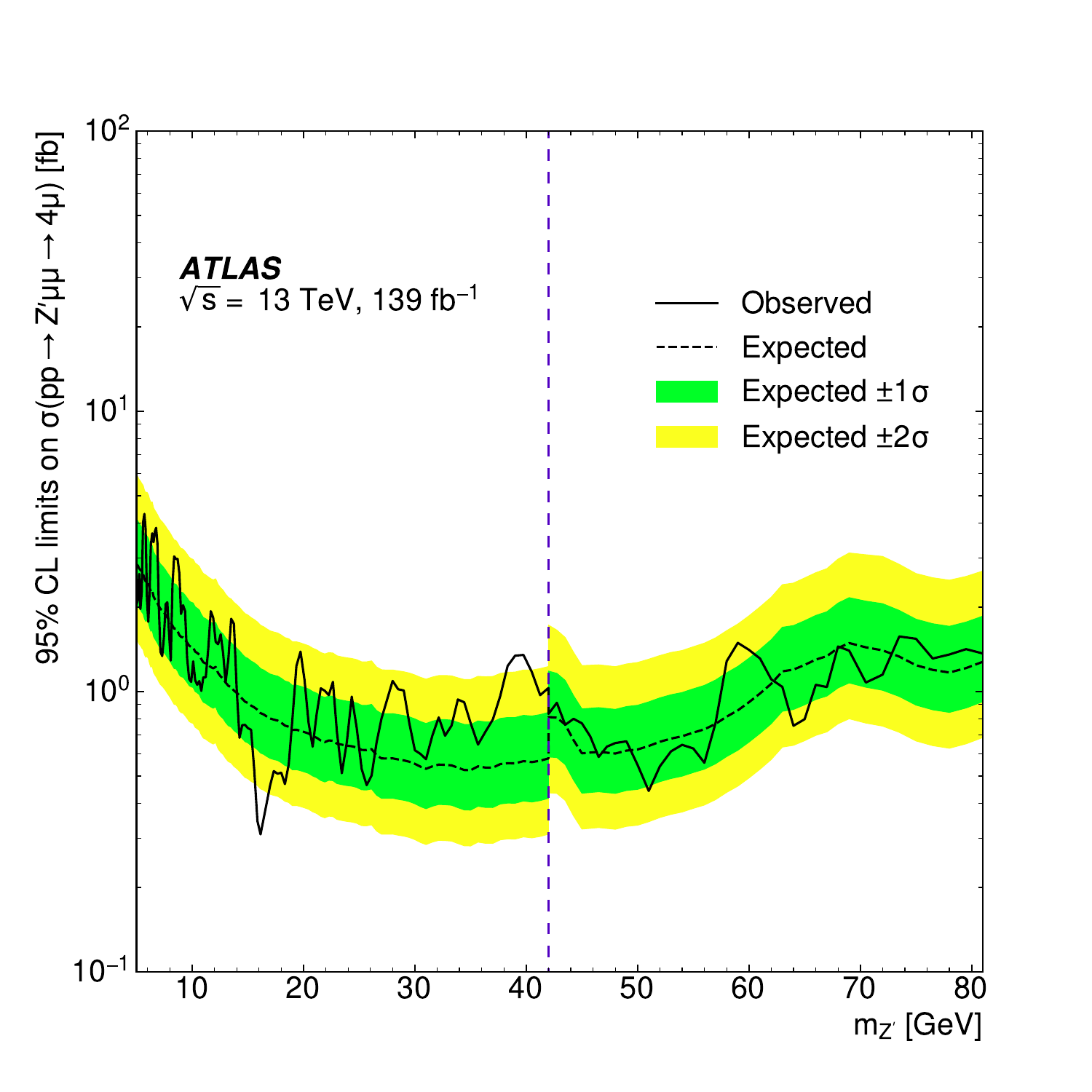}
\label{fig:xs_limit}
}
\subfloat[]{
\includegraphics[width=0.49\textwidth]{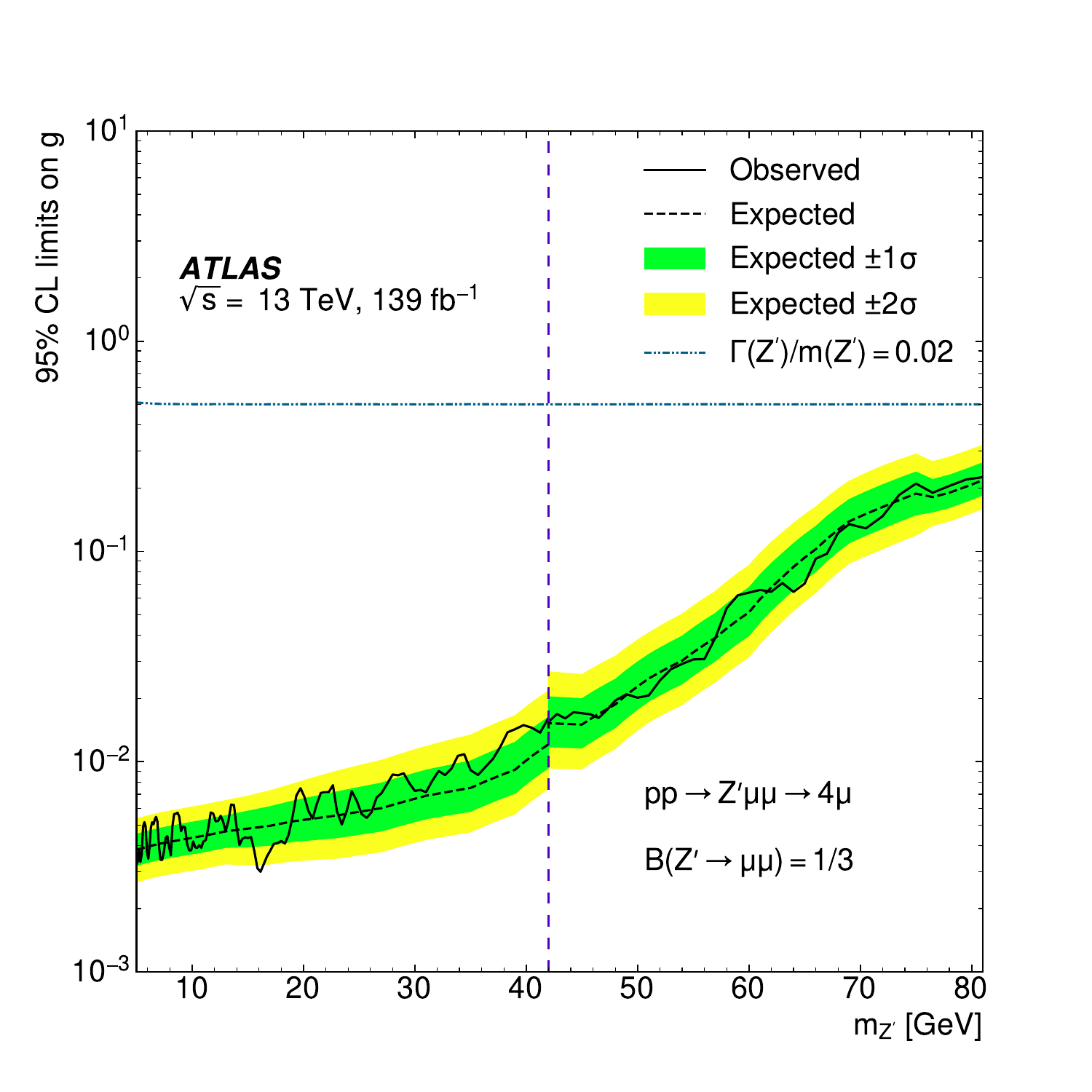}
\label{fig:coup_limit}
}
\caption{
95\% CL upper limits (expected and observed) on the cross-sections times branching fraction (a) and coupling parameter (b).
The discontinuity at 42 GeV represents the border of the low/high mass classifiers.
Considering the dimuon mass resolution of 2\% of the ATLAS detector, the horizontal dashed line in (b) indicates the upper limit on the valid coupling parameter of the model used in this analysis.
}
\label{fig:limit}
\end{figure}
 
Motivated by theoretical interpretations in \Refn{~\cite{Altmannshofer:2016jzy}},
a 2-dimensional exclusion contour at 95\% CL in the parameter-space of $(m_{Z'}, \coupzp)$ of the $L_\mu-L_\tau$ model from this analysis is produced and shown in \Fig{\ref{fig:limit_comp}}.
The parameter space exclusion regions calculated by theorists using data from
the Neutrino Trident experiment~\cite{PhysRevLett.66.3117}
and the $B_s$ mixing measurements by a global analysis performed in \Refn{~\cite{Altmannshofer:2016jzy}}
are also shown in \Fig{\ref{fig:limit_comp}}.
This had left a large gap in the parameter space not yet excluded.
This gap is now largely excluded by this analysis.

\begin{figure}[!htbp]
\centering
\includegraphics[width=0.6\textwidth]{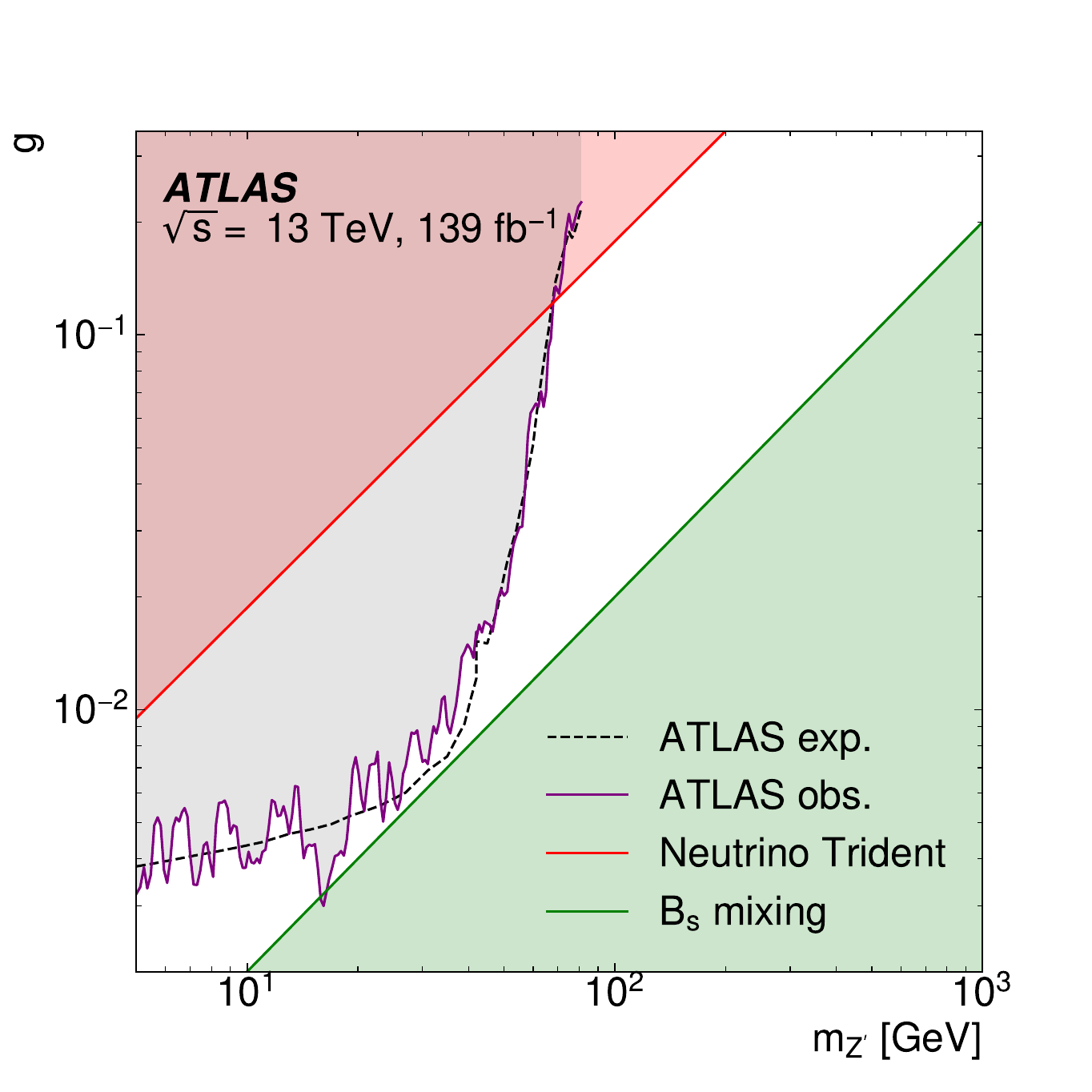}
\caption{
The coupling parameter $g$ limits from this search as a function of the $Z'$ mass
compared to the limits~\cite{Altmannshofer:2016jzy} from the Neutrino Trident (red) and the $B_s$ mixing (green) experimental results.
}
\label{fig:limit_comp}
\end{figure}


\FloatBarrier
 
\section{Conclusion}
\label{sec:conclusion}

A search for a new vector gauge boson $Z'$ predicted by the $L_\mu-L_\tau$ models has been performed with a 4$\mu$ final state in the invariant mass range of [80, 180]~\GeV{}, excluding the Higgs boson mass window [110, 130]~\GeV{}, using 139~\ifb{} of $\sqrt{s} = $ 13~\TeV{} proton-proton collision data collected with the ATLAS detector.
No significant excess of events over the expected SM background is observed.
Therefore, upper limits are set on the \Zp production cross-section times the decay branching fraction
of the $pp \rightarrow Z'\mu^+\mu^- \rightarrow \mu^+\mu^-\mu^+\mu^-$ process,
varying from 0.31 to 4.3 fb at 95\% CL,
in a $Z'$ mass range of [5, 81]~\GeV{},
from which the coupling strength \coupzp of the \Zp to muons above 0.003 to 0.2 (depending on the \Zp mass) are excluded in the same mass range.
This search shows significant sensitivity improvements over previous indirect and direct searches of the \Zp with 4$\mu$ final state.
An interesting parameter space of the $L_\mu-L_\tau$ model prediction that was not excluded by previous experiments
is now largely excluded by this search.


\clearpage
\section*{Acknowledgements}


We thank CERN for the very successful operation of the LHC, as well as the
support staff from our institutions without whom ATLAS could not be
operated efficiently.
 
We acknowledge the support of
ANPCyT, Argentina;
YerPhI, Armenia;
ARC, Australia;
BMWFW and FWF, Austria;
ANAS, Azerbaijan;
SSTC, Belarus;
CNPq and FAPESP, Brazil;
NSERC, NRC and CFI, Canada;
CERN;
ANID, Chile;
CAS, MOST and NSFC, China;
Minciencias, Colombia;
MEYS CR, Czech Republic;
DNRF and DNSRC, Denmark;
IN2P3-CNRS and CEA-DRF/IRFU, France;
SRNSFG, Georgia;
BMBF, HGF and MPG, Germany;
GSRI, Greece;
RGC and Hong Kong SAR, China;
ISF and Benoziyo Center, Israel;
INFN, Italy;
MEXT and JSPS, Japan;
CNRST, Morocco;
NWO, Netherlands;
RCN, Norway;
MEiN, Poland;
FCT, Portugal;
MNE/IFA, Romania;
JINR;
MES of Russia and NRC KI, Russian Federation;
MESTD, Serbia;
MSSR, Slovakia;
ARRS and MIZ\v{S}, Slovenia;
DSI/NRF, South Africa;
MICINN, Spain;
SRC and Wallenberg Foundation, Sweden;
SERI, SNSF and Cantons of Bern and Geneva, Switzerland;
MOST, Taiwan;
TENMAK, T\"urkiye;
STFC, United Kingdom;
DOE and NSF, United States of America.
In addition, individual groups and members have received support from
BCKDF, CANARIE, Compute Canada and CRC, Canada;
COST, ERC, ERDF, Horizon 2020 and Marie Sk{\l}odowska-Curie Actions, European Union;
Investissements d'Avenir Labex, Investissements d'Avenir Idex and ANR, France;
DFG and AvH Foundation, Germany;
Herakleitos, Thales and Aristeia programmes co-financed by EU-ESF and the Greek NSRF, Greece;
BSF-NSF and GIF, Israel;
Norwegian Financial Mechanism 2014-2021, Norway;
NCN and NAWA, Poland;
La Caixa Banking Foundation, CERCA Programme Generalitat de Catalunya and PROMETEO and GenT Programmes Generalitat Valenciana, Spain;
G\"{o}ran Gustafssons Stiftelse, Sweden;
The Royal Society and Leverhulme Trust, United Kingdom.
 
The crucial computing support from all WLCG partners is acknowledged gratefully, in particular from CERN, the ATLAS Tier-1 facilities at TRIUMF (Canada), NDGF (Denmark, Norway, Sweden), CC-IN2P3 (France), KIT/GridKA (Germany), INFN-CNAF (Italy), NL-T1 (Netherlands), PIC (Spain), ASGC (Taiwan), RAL (UK) and BNL (USA), the Tier-2 facilities worldwide and large non-WLCG resource providers. Major contributors of computing resources are listed in Ref.~\cite{ATL-SOFT-PUB-2021-003}.


\clearpage
\printbibliography
 
\clearpage
 
\begin{flushleft}
\hypersetup{urlcolor=black}
{\Large The ATLAS Collaboration}

\bigskip

\AtlasOrcid[0000-0002-6665-4934]{G.~Aad}$^\textrm{\scriptsize 101}$,
\AtlasOrcid[0000-0002-5888-2734]{B.~Abbott}$^\textrm{\scriptsize 119}$,
\AtlasOrcid[0000-0002-7248-3203]{D.C.~Abbott}$^\textrm{\scriptsize 102}$,
\AtlasOrcid[0000-0002-1002-1652]{K.~Abeling}$^\textrm{\scriptsize 55}$,
\AtlasOrcid[0000-0002-8496-9294]{S.H.~Abidi}$^\textrm{\scriptsize 29}$,
\AtlasOrcid[0000-0002-9987-2292]{A.~Aboulhorma}$^\textrm{\scriptsize 35e}$,
\AtlasOrcid[0000-0001-5329-6640]{H.~Abramowicz}$^\textrm{\scriptsize 150}$,
\AtlasOrcid[0000-0002-1599-2896]{H.~Abreu}$^\textrm{\scriptsize 149}$,
\AtlasOrcid[0000-0003-0403-3697]{Y.~Abulaiti}$^\textrm{\scriptsize 116}$,
\AtlasOrcid[0000-0003-0762-7204]{A.C.~Abusleme~Hoffman}$^\textrm{\scriptsize 136a}$
% ...
\bigskip
\\

% ...
$^{29}$Physics Department, Brookhaven National Laboratory, Upton NY; United States of America.\\
% ...
$^{35}$$^{(a)}$Facult\'e des Sciences Ain Chock, R\'eseau Universitaire de Physique des Hautes Energies - Universit\'e Hassan II, Casablanca;$^{(b)}$Facult\'{e} des Sciences, Universit\'{e} Ibn-Tofail, K\'{e}nitra;$^{(c)}$Facult\'e des Sciences Semlalia, Universit\'e Cadi Ayyad, LPHEA-Marrakech;$^{(d)}$LPMR, Facult\'e des Sciences, Universit\'e Mohamed Premier, Oujda;$^{(e)}$Facult\'e des sciences, Universit\'e Mohammed V, Rabat;$^{(f)}$Institute of Applied Physics, Mohammed VI Polytechnic University, Ben Guerir; Morocco.\\
% ...
$^{55}$II. Physikalisches Institut, Georg-August-Universit\"{a}t G\"ottingen, G\"ottingen; Germany.\\
% ...
$^{101}$CPPM, Aix-Marseille Universit\'e, CNRS/IN2P3, Marseille; France.\\
$^{102}$Department of Physics, University of Massachusetts, Amherst MA; United States of America.\\
% ...
$^{116}$Department of Physics, New York University, New York NY; United States of America.\\
% ...
$^{119}$Homer L. Dodge Department of Physics and Astronomy, University of Oklahoma, Norman OK; United States of America.\\
% ...
$^{136}$$^{(a)}$Departamento de F\'isica, Pontificia Universidad Cat\'olica de Chile, Santiago;$^{(b)}$Millennium Institute for Subatomic physics at high energy frontier (SAPHIR), Santiago;$^{(c)}$Instituto de Investigaci\'on Multidisciplinario en Ciencia y Tecnolog\'ia, y Departamento de F\'isica, Universidad de La Serena;$^{(d)}$Universidad Andres Bello, Department of Physics, Santiago;$^{(e)}$Instituto de Alta Investigaci\'on, Universidad de Tarapac\'a, Arica;$^{(f)}$Departamento de F\'isica, Universidad T\'ecnica Federico Santa Mar\'ia, Valpara\'iso; Chile.\\
% ...
$^{149}$Department of Physics, Technion, Israel Institute of Technology, Haifa; Israel.\\
$^{150}$Raymond and Beverly Sackler School of Physics and Astronomy, Tel Aviv University, Tel Aviv; Israel.\\
% ...
$^{a}$ Also Affiliated with an institute covered by a cooperation agreement with CERN.\\
$^{b}$ Also at An-Najah National University, Nablus; Palestine.\\
$^{c}$ Also at Borough of Manhattan Community College, City University of New York, New York NY; United States of America.\\
$^{d}$ Also at Bruno Kessler Foundation, Trento; Italy.\\
$^{e}$ Also at Center for High Energy Physics, Peking University; China.\\
$^{f}$ Also at Centro Studi e Ricerche Enrico Fermi; Italy.\\
% ...

\end{flushleft}


\end{document}